\documentclass[11pt,twoside,a4paper,cmspaper,final,collab]{cms-tdr}
\def\svnVersion{78dcd91}\def\svnDate{2026/09/16}\def\cmsCernNoTag{CERN-EP-2025-206}\def\cmsCernDate{\today}\def\cmsMessage{Published in the Journal of High Energy Physics as \href{http://dx.doi.org/10.1007/JHEP08(2026)049}{\doi{10.1007/JHEP08(2026)049}} and \href{https://doi.org/10.1007/JHEP10(2026)168}{\doi{0.1007/JHEP10(2026)168}}.}
\begin{document}\cmsNoteHeader{SMP-22-009}

\newcommand{\ptgamma}{\ensuremath{p_{T}^{\PGg}}\xspace}
\newcommand{\srprocess}{\ensuremath{\PZ(\to\PGn\PAGN)+\PGg}\xspace}

\newcommand{\zinv}{\ensuremath{\PZ\to\PGn\PAGn}\xspace}
\newcommand{\zinvg}{\ensuremath{\PZ(\PGn\PAGn)\PGg}\xspace}
\newcommand{\CP}{\ensuremath{CP}\xspace}
\providecommand{\PGgst}{{\HepParticle{\PGg}{}{\ast}}\xspace}
\providecommand{\PZst}{{\HepParticle{\PZ}{}{\ast}}\xspace}
\providecommand{\PGnell}{{\HepParticle{\PGn}{\Pell}{}}\xspace}
\providecommand{\PAGnell}{{\HepAntiParticle{\PGn}{\Pell}{}}\xspace}
\newlength\cmsTabSkip\setlength{\cmsTabSkip}{1ex}

\cmsNoteHeader{SMP-22-009}

\title{Measurement of the \texorpdfstring{$\PZ\PGg$}{Zgamma} production cross section and search for anomalous neutral triple gauge couplings in $\Pp\Pp$ collisions at \texorpdfstring{$\sqrt{s}=13\TeV$}{sqrt(s) = 13 TeV}}

\date{\today}

\abstract{
A measurement of the fiducial cross section of the associated production of a $\PZ$ boson and a high-\pt photon, where the \PZ decays to two neutrinos, and a search for anomalous triple gauge couplings are reported. The results are based on data collected by the CMS experiment at the LHC in proton-proton collisions at $\sqrt{s}=13\TeV$ during 2016--2018, corresponding to an integrated luminosity of 138\fbinv. The fiducial $\PZ\PGg$ cross section, where a photon with a \pt greater than 225\GeV is produced in association with a \PZ, and the \PZ decays to a $\PGn\PAGn$ pair ($\PZ(\PGn\PAGn)\PGg$), is measured to be $23.3^{+1.4}_{-1.3}\unit{fb}$, in agreement, within uncertainties, with the standard model prediction. The differential cross section as a function of the photon \pt has been measured and compared with standard model predictions computed at next-to-leading and at next-to-next-to-leading order in perturbative quantum chromodynamics. Constraints have been placed on the presence of anomalous couplings that affect the $\PZ\PZ\PGg$ and $\PZ\PGg\PGg$ vertex using the \pt spectrum of the photons. The observed 95\% confidence level intervals for \CP-conserving $h_3^{\PGg}$ and $h_4^{\PGg}$ are determined to be ($-$3.4, 3.5)$\times10^{-4}$ and ($-$6.8, 6.8)$\times10^{-7}$, and for $h_3^{\PZ}$ and $h_4^{\PZ}$ they are ($-$2.2, 2.2)$\times10^{-4}$ and ($-$4.1, 4.2)$\times10^{-7}$, respectively. These are the strictest limits to date on $h_3^{\PGg}$, $h_3^{\PZ}$ and $h_4^{\PZ}$. 
}

\hypersetup{%
pdfauthor={CMS Collaboration},%
pdftitle={Measurement of the Z  gamma production cross section and  search for anomalous neutral triple gauge couplings in pp collisions at sqrt(s) = 13 TeV},%
pdfsubject={CMS},%
pdfkeywords={CMS, aNTGC, monophoton}
}

\maketitle 
\section{Introduction} 
Measurements of the production cross sections of electroweak (EW) dibosons at collider experiments provide important tests of the electroweak sector of the standard model (SM). 
In the SM, vertices with three neutral electroweak bosons ($\PZ\PZ\PZ$, $\PZ\PGg\PGg$, and $\PZ\PZ\PGg$) are not allowed at the tree level, as they violate the $SU(2)_\mathrm{L} \times U(1)_\mathrm{Y}$ symmetry. However, in the presence of new physics, anomalous neutral triple gauge couplings (aNTGCs) could arise that can be introduced as extension to the SM using the Lagrangian terms of all possible Lorentz structures, or in an effective field theory, where the operators have dimensions higher than those of the SM electroweak Lagrangian~\cite{antgc_hagiwara}.

\begin{figure}[ht]
        \centering
        \includegraphics[width=0.5\linewidth]{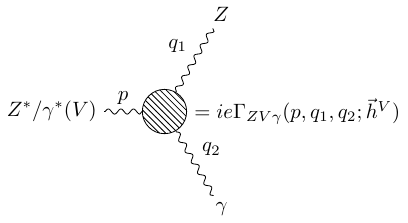}
        \caption{The Feynman diagram for the $\PZ\PV\PGg$ vertex, where $\PV~=\PGg^{*}/\PZ^{*}$, in the presence of an aNTGC.}%
        \label{fig:antgc_vertex}
\end{figure}

To estimate these effects, vertex functions have been calculated in Refs.~\cite{antgc_hagiwara, antgc_baur1, antgc_baur2} to describe the most general Lorentz- and $U_\mathrm{Y}(1)$-invariant interactions of an incoming off-shell \PZ boson or photon, and an outgoing on-shell $\PZ$ with a photon (Fig.~\ref{fig:antgc_vertex}). The strength of this interaction is determined by four parameters: $h_1^\PV$ and $h_2^\PV$, which are \CP violating, and $h_3^\PV$ and $h_4^\PV$, which are \CP conserving. As an example, the $\PZ\PZ\PGg$ vertex function (in the case of outgoing on-shell $\PZ$/$\PGg$) is defined as:
\begin{multline}
        \label{eq:lagrangian}
            \Gamma^{\alpha\beta\mu}_{\PZ\PZ\PGg} = \frac{p^2 - q_1^2}{m_\PZ^2} \left[h_1^\PZ(q_2^\mu g^{\alpha\beta} - q_2^\alpha g^{\mu\beta})+\frac{h_2^\PZ}{m_\PZ^2}p^\alpha\Big[ (p\cdot q_2)g^{\mu\beta} - q_2^\mu p^\beta\Big]+\right.\\
            \left. h_3^\PZ\epsilon^{\mu\alpha\beta\rho}q_{2\rho}+ \frac{h_4^\PZ}{m_\PZ^2}p^{\alpha}\epsilon^{\mu\beta\rho\sigma}p_{\rho}q_{2\sigma}\right],\\
\end{multline}
where $\Gamma^{\alpha\beta\mu}_{Z\PGg\PGg}$ can be obtained with the replacements:
        \begin{equation*}
        \frac{p^2 - q_1^2}{m_\PZ^2}\to \frac{p^2}{m_\PZ^2}
        \text{ and } h_i^\PZ \to h_i^\PGg.
        \end{equation*}
Here, $g^{\alpha\beta}$ denotes the Minkowski metric tensor with signature $(+,-,-,-)$, and 
$\epsilon^{\mu\alpha\beta\rho}$ is the totally antisymmetric Levi-Civita tensor, defined such that 
$\epsilon_{0123} = +1$, and repeated indices are summed over the four space-time coordinates $(0,1,2,3)$.        

\begin{figure}[tb]
	\centering
    \includegraphics[width=0.35\linewidth]{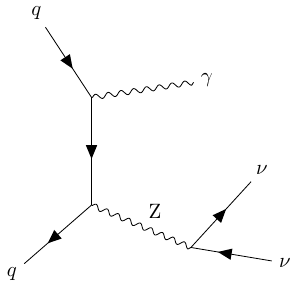}
    \includegraphics[width=0.4\linewidth]{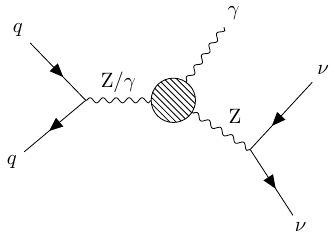}
    \caption{The leading-order Feynman diagram for SM $\PZ\PGg$ production in the $t$-channel (left), and (right) the additional diagram for the beyond SM \textit{s}-channel  $\PZ\PGg$ production due to the presence of aNTGCs.}
    \label{fig:antgc_s_t_channel}
\end{figure}

 The existence of such an effective vertex can be detected as either an excess, or as a deficit relative to the SM prediction in the kinematic distributions of the $\PZ\PGg$ production, since the aNTGC introduces an additional channel for this process (Fig.~\ref{fig:antgc_s_t_channel}), changing the cross section to $\sigma_\text{total}=\sigma_\text{SM} + \sigma_\text{interference} + \sigma_\text{aNTGC}$~\cite{antgc_baur2}.
 In the presence of aNTGCs, the total production cross section of the $\PZ\PGg$ process depends on the strength of the couplings. The leading-order corrections to the SM cross-section scale linearly with the \CP-conserving couplings due to the interference between the SM and aNTGC amplitudes, while second-order corrections arise from terms quadratic in these strengths.
In contrast, the \CP-violating couplings do not interfere with either the SM amplitude or an anomalous \CP-conserving amplitude, and  their impact on the Lagrangian is manifested only at the quadratic level~\cite{antgc_gounaris_1}. 
However, when the outgoing photon has a high transverse momentum, 
the interference term becomes negligible and
the Lagrangian associated with the \CP-conserving couplings becomes, as it is for the \CP-violating couplings, quadratically sensitive~\cite{antgcInterferenceBellazzini}.  
At energies exceeding a few hundred \GeVns, the limits on $h_1^\PV$ and $h_2^\PV$ become comparable to those on $h_3^\PV$ and $h_4^\PV$, respectively, and
consequently in this paper only a search of the \CP-conserving couplings $h_3^\PV$ and $h_4^\PV$ is reported.

In the analysis presented here, we have used the vertex function approach, to enable comparison with previous results. While not considered in this paper, a new formalism that includes contributions
from off-shell decays $\PZ^{*} \rightarrow \PGn\PAGn$ has been presented in~\cite{Ellis:2022zdw}, and a method to compare the two frameworks can be found in~\cite{CMS:2025cxp}.

The neutral decay of the \PZ boson ($\zinv$) is the most suitable channel to detect aNTGCs, as it provides a clean final state signature with a 20\% branching fraction, compared to the charged lepton decay channel where it is 10\%. The hadronic decay channel of the \PZ boson has the highest branching fraction of 70\%, but also has a much larger background from $\PGg+$jets events. 

Measurements of the production cross section of $\zinvg$ in hadron collisions and studies of the triple $\PZ\PZ\PGg$ and $\PZ\PGg\PGg$ vertices  were first performed by the D0 Collaboration~\cite{d0_ZG_study_2005}, and subsequently by the CDF~\cite{cdfZGmeasurement2010,cdfAntgcSearch2011}, CMS~\cite{CMS_8TeV_2,CMS_8TeV_1}, and ATLAS~\cite{atlas_8TeV,atlas_13TeV} Collaborations. In addition, limits on the anomalous couplings have been set in the charged-lepton final state 
$\PZ(\Pellp \Pellm)\PGg$, where $\Pell = \Pe$, \PGm, by the D0, CDF, ATLAS, and CMS Collaborations~\cite{d0_ZG_study_2005, d0_ZG_llg_study_2007,cdfZGmeasurement2010, cdfAntgcSearch2011, atlas_7tev_llg, CMS_8TeV_1}.

We report here a search for aNTGCs ($\PZ\PZ\PGg$ and $\PZ\PGg\PGg$)
with a high-\pt photon and a large missing momentum and a measurement of the production cross section of the $\PZ\PGg$ process in data collected by the CMS experiment in proton-proton (pp) collisions at $\sqrt{s}=13\TeV$ during 2016--2018, corresponding to an integrated luminosity of 138\fbinv.

In this paper the CMS detector is described in Section~\ref{sec:CMS} and the analysis strategy in Section~\ref{sec:anaStrat}. The data sets and the simulated samples used to model the signal and the backgrounds are described in Section~\ref{sec:datasets}. The criteria used to select physics objects and to define the signal and control regions are described in Sections~\ref{sec:photonID} and~\ref{sec:event_selection}, and in Section~\ref{sec:bkgEst} the methods used to estimate the backgrounds are discussed. The procedures used to estimate the systematic uncertainties that affect the modeling are described in Section~\ref{sec:unc}. The  methods used to estimate the overall uncertainty of the obtained results are described in Sections~\ref{sec:statAn} and~\ref{sec:res}. Tabulated results are provided in the HEPData record for this analysis~\cite{hepdata}.

\section{The CMS detector and object reconstruction}
\label{sec:CMS}
The CMS apparatus~\cite{CMS:2008xjf,CMS:2023gfb} is a multipurpose, nearly hermetic detector, designed to trigger on~\cite{CMS:2020cmk,CMS:2016ngn,CMS:2024aqx} and identify electrons, muons, photons, and (charged and neutral) hadrons~\cite{CMS:2020uim,CMS:2018rym,CMS:2014pgm}. 
The central feature of the CMS apparatus is a superconducting solenoid of 6\unit{m} internal diameter, providing a magnetic field of 3.8\unit{T}. 
A silicon pixel and strip tracker, a lead tungstate crystal electromagnetic calorimeter (ECAL) with a silicon strip preshower detector (ES) placed at the front faces of the two endcap calorimeters, and a brass and scintillator hadron calorimeter (HCAL), each composed of a barrel and two endcap sections, are installed inside the solenoid. Avalanche photodiodes (APDs) are used to readout the barrel section of the ECAL (EB), while vacuum photo-triodes are used to readout the ECAL endcaps (EE). Forward calorimeters extend the $\eta$ coverage 
provided by the barrel and endcap detectors, where $\eta$ is the pseudorapidity and is defined using the polar angle $\theta$ as $\eta = -\ln[\tan(\theta/2)]$. Muons are detected in gas-ionization chambers embedded in the steel flux-return yoke outside the solenoid. 
A more detailed description of the CMS detector, together with a definition of the coordinate system and the relevant kinematic variables, 
can be found in Ref.~\cite{CMS:2008xjf}.

Events of interest are selected using a two-tier trigger system. The first level, composed of custom hardware processors, uses information from
the calorimeters and muon detectors to select events at a rate of around 100\unit{kHz} within a fixed latency of about 4\mus~\cite{CMS:2020cmk}. 
The second level, known as the high-level trigger, consists of a farm of processors running a version of the full event reconstruction software 
optimized for fast processing and reduces the event rate to around 1\unit{kHz} before data storage~\cite{CMS:2016ngn, CMS:2024aqx}.

The global event reconstruction (also called particle-flow event reconstruction~\cite{CMS:2017yfk}) aims to reconstruct and identify each individual particle in an event, with an optimized combination of all subdetector information. In this process, the identification of the particle type (photon, electron, muon, charged hadron, neutral hadron) plays an important role in the determination of the particle direction and energy. Photons (\eg coming from \PGpz decays or from electron bremsstrahlung) are identified as ECAL energy clusters not linked to the extrapolation of any charged particle trajectory to the ECAL. Electrons (\eg coming from photon conversions in the tracker material or from \PB hadron semileptonic decays) are identified as a primary charged particle track and potentially many ECAL energy clusters corresponding to this track extrapolation to the ECAL and to possible bremsstrahlung photons emitted along the way through the tracker material. Muons (\eg from \PB hadron semileptonic decays) are identified as tracks in the central tracker consistent with either a track or several hits in the muon system, and associated with calorimeter deposits compatible with the muon hypothesis. Charged hadrons are identified as charged particle tracks neither identified as electrons, nor as muons. Finally, neutral hadrons are identified as HCAL energy clusters not linked to any charged hadron trajectory, or as a combined ECAL and HCAL energy excess with respect to the expected charged hadron energy deposit.
The primary vertex (PV) is taken to be the vertex corresponding to the hardest scattering in the event, evaluated using tracking information alone, as described in Section 9.4.1 of Ref.~\cite{CMS-TDR-15-02}.

Hadronic jets are clustered from all the PF candidates in an event using the infrared and collinear-safe anti-\kt
algorithms~\cite{Cacciari:2008gp, Cacciari:2011ma} with a distance parameter of 0.4. Jet momentum is determined as the vector sum of all particle momenta
in the jet. From simulation, this is found to be on average within 5 to 10\% of the true momentum over the whole \pt spectrum and detector acceptance.
Additional $\Pp\Pp$ interactions within the same, or nearby bunch crossings (pileup) can contribute energy depositions in the calorimeter and additional tracks,
increasing the apparent jet momentum. To mitigate this effect, tracks identified as originating from pileup vertices are discarded
and a correction is applied to account for any residual contributions~\cite{Sirunyan:2020foa}.

Corrections to the jet energy are derived from simulation studies and applied so that the average measured jet energy is identical to particle level jets.
\textit{In situ} measurements of the momentum balance in dijet, $\PGg+$jet, $\PZ+$jet, and multijet events are used to determine any residual differences
between the jet energy scale in the observed events and in simulation, and corrections factors are applied~\cite{CMS:2016lmd}.

The missing transverse momentum vector \ptvecmiss is defined as the projection onto the plane perpendicular to the beam axis of the negative vector sum of the momenta of all reconstructed particle-flow objects in an event. Its magnitude is referred to as \ptmiss~\cite{Sirunyan:2019kia}.
The \ptvecmiss vector is modified to account for corrections to the energy scale of the reconstructed jets.

For leptons, corrections are applied as weights to the simulated events to account for differences between data and simulation in the muon and electron reconstruction efficiencies. 
These corrections are determined using the ``tag-and-probe'' technique~\cite{Khachatryan:2010xn} with $\PZ \to \Pep\Pem$~events.
Differences between the muon momentum scale in data and simulation are corrected for using a calibration derived
in $\PZ\to \PGmp\PGmm$ events~\cite{Bodek:2012id,CMS:2018rym}.
An additional resolution smearing is included in the simulation to match the resolution in data.
Similarly, scale and resolution corrections are applied to simulated electrons and photons to better match the energy scale and resolution measured in the data.

Spurious high-\ptmiss events can be due to a variety of reconstruction failures, detector malfunctions, or non-collision backgrounds.
Such events are rejected by event filters designed to identify more than 85\% of the spurious high-\ptmiss events, with a mistagging rate of less than 0.1\%~\cite{Sirunyan:2019kia}.

\section{Analysis strategy} 
\label{sec:anaStrat}

\begin{figure}[ht]
    \centering
      \includegraphics[width=0.55\linewidth]{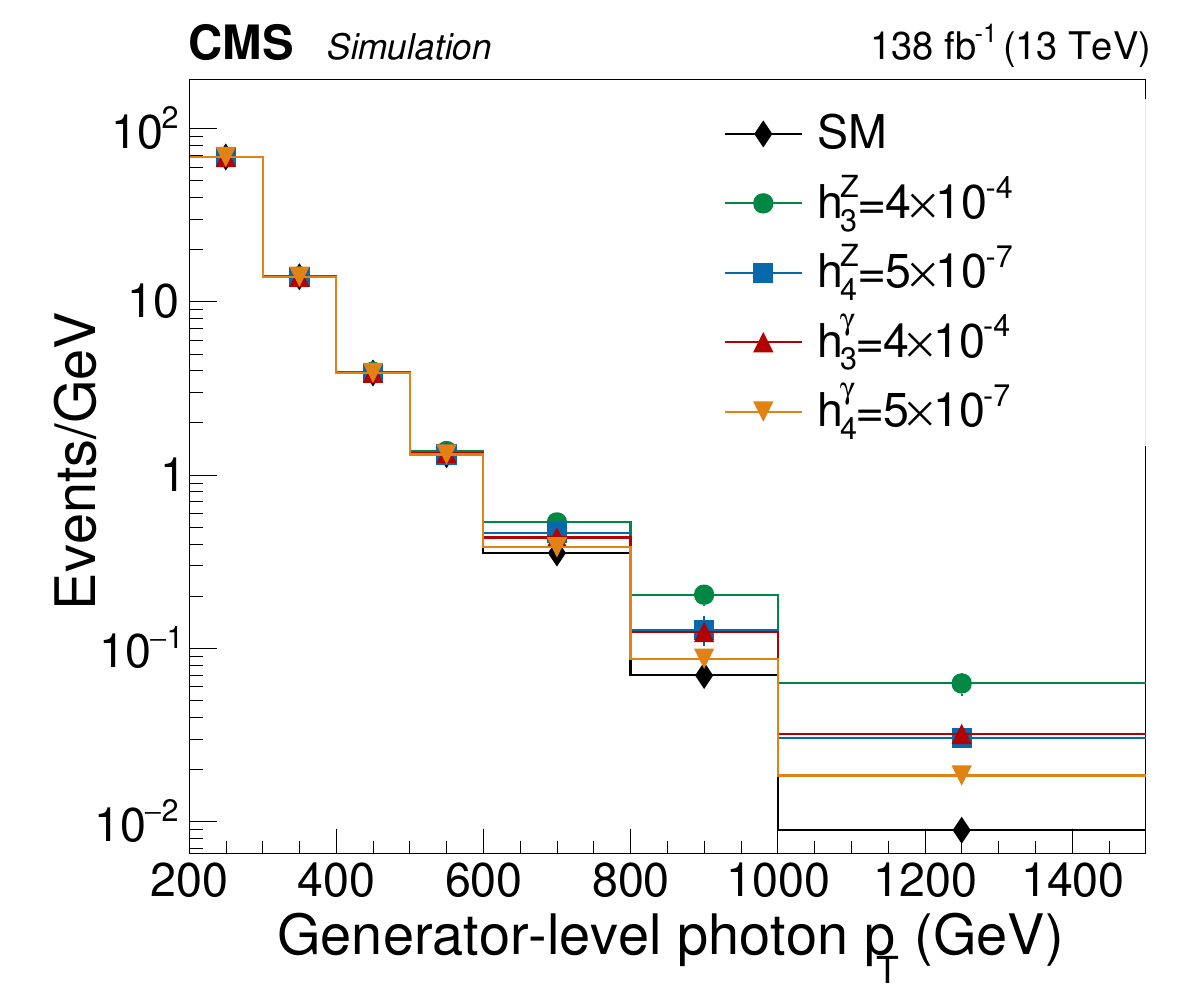}
	\caption{The generator-level photon \pt spectrum for various values of aNTGC couplings and the SM predictions.} 
      \label{fig:aNTGC_excess} 
\end{figure}

The signature of $\PZ(\PGn\PAGn)\PGg$ events is a high-$\pt$ photon associated with large $\ptmiss$. Figure~\ref{fig:aNTGC_excess} shows the generator-level photon \pt spectrum for different values of the anomalous coupling parameters. 
The sensitivity to aNTGCs increases with the photon \pt, with the greatest sensitivity above \pt$\approx 500\GeV$. To better constrain all the SM background contributions to this process, we require events to have a photon with a \pt greater than 200\GeV. 
To select these high-$\pt$ photons and to distinguish them from jets a dedicated identification algorithm, described in Section~\ref{sec:photonID}, has been specifically developed for this analysis. 

Backgrounds to the $\PZ(\PGn\PAGn)\PGg$ process come from two main sources: one is the process $\PW \to \Pe\PAGne$, where the electron is reconstructed as a photon due to an inefficiency of the pixel tracker that contributes approximately 43\% to the total background, and 
the $\PW(\Pell\PAGn_{\Pell})\PGg$ process, where the lepton is not identified, contributes around 33\% of the total background. 
Both these background sources are 
estimated from data using control samples. 
Other smaller backgrounds are events with a photon coming from the processes $\PZ/\PGg^{\ast}(\Pellp\Pellm)\PGg$, $\ttbar\PGg$, $\PW\PV\PGg$, and $\PQt$+$\PGg$, or with a final-state radiation photon in diboson or $\PW$ production. The contributions from these processes are estimated using simulation to be 16\% to the total background. 
An additional background contribution, where a jet is misidentified as a photon in association with large $\ptmiss$, is estimated from data to be 4\% of the total background.

Direct ionization of the APDs~\cite{spikes_petyt_mitigation_2012} in EB, spikes, and bremsstrahlung radiation from beam-halo muons~\cite{beamHaloSingh2010}  in both EB and EE can mimic high-\PT photons and contribute approximately 4\% to the total background.  

\section{Dataset and simulation} 
\label{sec:datasets}
Samples of simulated data are used to model the background processes and to 
estimate the SM and aNTGC signals. 
These samples are produced using \MGvATNLO~\cite{MadGraph} v2.6.5 at QCD and EW leading order (LO), 
with up to 2 additional jets using the CP5~\cite{Sirunyan:2019dfx} underlying event \PYTHIA v8.240~\cite{pythia8} tune. 
\PYTHIA is used for fragmentation and hadronization and 
the detector response is simulated with the \GEANTfour package~\cite{GEANT4}.

Correction factors ($K$-factors) that depend on the \pt of the photon are applied to account for next-to-next-to-leading order (NNLO) in QCD~\cite{kFacQCD_Grazzini} and at next-to-leading order (NLO) in EW theory~\cite{kFacEWZG_Denner,kFacEWWG_Denner}. In these references, the NLO EW K-factors for
$\PZ(\PGn\PAGn)\PGg$ and $\PW(\Pell \PAGn_\Pell)\PGg$ 
were calculated for $\sqrt{s}=14\TeV$, 
and have been recalculated for $\sqrt{s}=13\TeV$ using the NNPDF31luxQED~\cite{Bertone:2017bme} PDF set. 
For the SM $\zinvg$ sample events where the photon \pt exceeds 200\GeV, 
at $\sqrt{s}=13\TeV$, the QCD $K$-factors decrease from 1.55 to 1.2, 
while the EW $K$-factors decrease from  0.95 to 0.3 for the \pt range of 200--2000\GeV. 
For the $\PW(\Pell\PGnell)\PGg$ samples the QCD $K$-factors decrease from  1.5 to 1.2, while the EW $K$-factors 
decrease from 0.95 to 0.4 for the \pt range of 200--2000\GeV. 
Pileup events are simulated and added to the hard scattering event. For each data-taking year the simulated events are weighted so that the number of pileup events matches the rate observed in the data.

Events from the standard model processes ${\PZ (\Pellp\Pellm)} \PGg$, $\PGg+\text{jets}$, $\PGg\PGg$, $\ttbar\PGg$, and $\PQt\PGg$ are produced at NLO in QCD using \MGvATNLO. 
Other background processes: $\PW\PW$, $\PW\PZ$, $\PZ\PZ$, $\PW (\PGm\PAGnGm)$, and $\PW (\PGt\PAGnGt)$ that have a very small contribution in the analysis were generated at LO with \PYTHIA8.

\section{Photon identification} 
\label{sec:photonID}

A dedicated high-$\pt$ photon identifier (ID)~\cite{WadudMohammadAbrar2023Motp} is used to distinguish high-\pt photons from jets. It uses a combination of the output of a boosted decision tree (BDT) and selections on the shower shape variables and the \pt and $\abs{\eta}$ of the photon. The ID was trained using simulated events with high-\pt photons and with a simulated QCD sample to model the background that was generated with \MGvATNLO. Photons are identified by their BDT score and selections on three additional variables, which are the ratio of energies in the HCAL to the ECAL, the degree of isolation in the ECAL, and the degree of isolation in the tracker, determined after corrections for pileup and clustering inefficiencies have been applied.  
The efficiency of the ID for photons between 200 and 2000\GeV is estimated to be $(92 \pm 1)\%$, where the uncertainty is estimated from the variation of efficiency with \pt. This photon ID has a higher efficiency for photons with \pt greater than 200\GeV and a lower fake rate than the standard photon ID used in CMS, which is optimized for lower values of \pt~\cite{CMS:2020uim}. Corrections applied as scale factors to account for the mismodeling of the identification efficiency are derived as functions of $\eta$ and \pt using the tag-and-probe technique with $\PZ\to \Pep\Pem$ events, where the electrons are reconstructed as photons. The differences between the efficiencies found in data and simulation are typically less than 4\%, with none greater than 10\%. 

\section{Event selection} 
\label{sec:event_selection}
The trigger system selected candidate events where the \pt of the photon ($\pt^{\PGg}$) was above a threshold of 175\GeV in 2016, and 200\GeV in 2017 and 2018. 
This change was due to an increase in the instantaneous luminosity after 2016 requiring the trigger threshold to be raised to maintain the trigger rate within the bandwidth limits. In addition to the requirement on $\pt^{\PGg}$, the ratio between the energy deposited in the HCAL and in the ECAL associated with the photon candidate was required to be less than 0.15 for photons in the EB and less than 0.1 for photons in the EE.

During the 2016 and 2017 data-taking there was a gradual shift in the timing of the inputs of the ECAL L1 trigger in the region $2.5 < \abs{\eta} < 3.0$ 
causing the trigger to be inefficient. For events containing a photon or an electron with $\pt>50\GeV$, or a jet with $\pt>100\GeV$
in this region, there was an efficiency loss of $\approx$10--20\%, depending on \pt, $\eta$, and when the data were taken. 
This resulted in a reduction of $\approx$2\% in the number of events used in the analysis, correction factors for this effect were computed from data and applied to the event acceptance.

Signal events are required to have $\ptmiss>200\GeV$ and one photon with $\pt^{\PGg}>225\GeV$, to be well above the trigger \pt threshold. 
The $\abs{\eta^{\PGg}}$ of the photon candidate is required not to be between 1.44 and 1.57~\cite{CMS:2020uim} to avoid the region between the EB and the EE, and to have no more than one associated hit in the pixel detector~\cite{CMS:2020uim}.

To suppress beam-halo muons that can mimic high-\pt photons when they traverse the EB, the requirement $\mathrm{MIP}_{\text{tot}}<4.9\GeV$ is applied to photons detected in the EB, where $\mathrm{MIP}_{\text{tot}}$ is the maximum of the total calorimeter energy summed along all possible paths of beam-halo that pass through the cluster. 

Signals from spikes are suppressed using the method described in Ref.~\cite{spikes_petyt_mitigation_2012}. To further suppress this background additional requirements are applied to the events. These are that the signal in the ECAL is within 3\unit{ns} of the bunch crossing time,
and that the ratio of energy of the seed crystal to the higher of two neighboring crystals in the $\eta$ direction is $ > 0.01$. 

To remove events with a misreconstructed $\ptmiss$ due to the jet energy resolution, events are discarded where at least one jet has a transverse momentum greater than 30\GeV  
and the angle $\Delta \phi$ between the jet and \ptvecmiss is less than 0.5.
Events are also required to have $\pt^{\PGg}/\ptmiss$ less than 1.4 to reject $\PGg+$jet background events.  
The value of 1.4 was selected to optimize the suppression of background events, in particular those from $\PGg$ + jet events. These two criteria remove more than 99\% of the $\PGg+$jet background events. 
The selected photon and the missing transverse momentum are also required to be separated in azimuthal angle $\Delta\phi(\pt^{\PGg}, \ptmiss) > 2.0$ to suppress $\PW (\Pe \PAGne)$ events where the electron has one, or no hits in the pixel detector.

During the 2018 data taking, two modules of the HCAL were inoperable, affecting the jet identification and jet energy measurement in the region $-1.57 < \phi < - 0.87$ and $-3.0 < \eta < -1.3$.
Events with a jet with $\pt^\text{jet}>15\GeV$ in this region are excluded. 
The selected photon candidate is also required to be outside of this region to suppress backgrounds from jets misidentified as photons.  
To suppress background events from $W(\Pell\PAGnell)\PGg$ processes, events with an electron or a muon with $\pt>10\GeV$ and passing loose identification criteria~\cite{CMS:2020uim, CMS-DP-2020-040} are also discarded. Table~\ref{tab:evSel} summarizes the principal selections used in the signal region (SR). 

Two control regions, $\text{CR}_\mathrm{e}$ and $\text{CR}_{\mu}$, are used to estimate the $W(\Pell\PAGnell)\PGg$ background. 
Events in these CRs are required to have an electron or a muon candidate, identified with tight criteria, with $\pt>30\GeV$, a photon satisfying the same signal requirements as in the SR, and the transverse mass of the $\ptmiss + \ell$ system less than 160\GeV. For these events the vector sum $\vec{R}_\mathrm{T} = \ptvecmiss+ \vec\pt^{\ell}$ is a proxy for the $\ptvecmiss$ in the SR, and the same requirements applied on \ptvecmiss in the SR are applied on $R_\mathrm{T}$. 
The selections applied in the SR to account for the inoperable HCAL modules in 2018 are applied with the additional requirement that the lepton is not in this region. Events are not included in the CRs if they contain a jet that has a $\pt^\text{jet}>30\GeV$ and an angular separation from $\vec{R}_\mathrm{T}$, $\Delta \phi(\text{jet}, \vec{R}_\mathrm{T})<0.5$. We also only include leptons that have $\Delta R = \sqrt{\smash[b]{(\Delta\eta)^2+(\Delta\phi)^2}}$
between the lepton and the jet greater than 0.5. 
Additionally, the lepton and the photon must be separated by $\Delta R (\Pe, \PGg)>0.5$. 

\begin{table}
\centering
\renewcommand{\arraystretch}{1.3}
\caption{Principal signal event selection criteria. Further details of the event selection, including fiducial criteria, are provided in the text.}
\begin{tabular}{p{0.6cm} p{13.4cm}}
\hline
\multicolumn{2}{l}{Events are selected if the event contains:}\\
& Exactly one photon with $p_T^{\PGg}>225$ GeV, with\\
& $H/E<0.15 (0.1)$ EB, (EE); $(\abs{\eta^{\PGg}}<1.44 )\cup  (1.57 < \abs{\eta^{\PGg}} < 2.5)$; \\
&$\ptmiss>200$ GeV; $p_T^{\PGg}/\ptmiss < 1.4$,
and $\Delta\phi(\ptvecmiss,\PGg)>2$. \\

\multicolumn{2}{l}{Events are not selected if the event contains:} \\
& a jet with $\pt^{\text{jet}}>30$ GeV and $\Delta\phi(\text{jet}, \ptvecmiss) < 0.5$, \\
& or a muon with  $\pt^{\Pgm} > 10$ GeV, \\ &or 
an electron with $\pt^\Pe > 10$ GeV and
 $\Delta R(\Pe,\gamma)>0.1$. \\
\hline
\end{tabular}
\label{tab:evSel}
\end{table}

\section{Background estimation}\label{sec:backgrounds}
\label{sec:bkgEst}

\subsection{Jets misidentified as photons}
\label{sec:jet_fakes}
Electromagnetic deposits in the ECAL produced by jets with few tracks and a small signal in the HCAL can be misidentified as photons, and,
when associated with a high $\ptmiss$ are a background to the $\PZ(\PGn\PAGn)\PGg$ signal. This background, which comes mainly from the $\PZ(\PGn\PAGn)+\text{jets}$ process and from events with multiple jets, is estimated by first computing the probability of misidentifying a jet in events with $\ptmiss<40\GeV$. 
For this, photon-like objects from the electromagnetic deposits of a jet were selected by inverting the isolation selection criteria. 
The misidentification ratio is then defined as the ratio of the observed number of photons that pass the photon ID to the number that pass less stringent photon ID requirements.  

Since the numerator can still be contaminated with genuine photons, we estimate this contamination by fitting the BDT score distribution in data to a combination of two templates, one for genuine photons and another for misidentified photons. The template for genuine photons is obtained from simulated $\Pp\Pp \to \PGg$+jets events and the one for misidentified photons is obtained directly from a data sideband, chosen by requiring the photon to have a large number of tracks within the photon's isolation cone. 

The misidentification ratio is then estimated in bins of $\pt^{\PGg}$ for the EB and EE separately. 
It is then applied to a sample of events in data with photon candidates that have passed the same less stringent criteria used for the denominator of the misidentification ratio to estimate the contribution of jets misidentified as photons in the SR.

The sources of uncertainty in the estimate of this background
are the definition of the background templates,
the criteria on the $\ptmiss$ for the ratio estimation, 
the modeling of the BDT variable for genuine photons in simulation, and the
binning and statistical uncertainties of the simulated signal photons in the template fitting. How these are taken into account is discussed in Section~\ref{sec:statAn}.

\subsection{Electrons misidentified as photons}\label{sec:lep_fakes}

Standard model processes, such as $\PW(\Pe\PAGne)$ and 
$\PW(\Pe\PAGne)\PZ(\PGn\PAGn)$ production, 
can mimic the $\PZ(\PGn\PAGn)\PGg$ signature when the electron is reconstructed as a photon, which occurs when one or more layer of the pixel detector does not detect the electron. 
The estimate of the electron-to-photon misidentification ratio is made with $\PZ\to \Pep\Pem$ events using the tag-and-probe method~\cite{CMS:2020uim}, with the electron tag required to have $\pt > 20\GeV$ and the probe to have $\pt> 200\GeV$. The invariant mass of the tag-probe pair is required to be between 60 and 120\GeV and a passing (failing) probe is then defined if the electron is (is not) reconstructed as an electron candidate~\cite{CMS:2020uim}. The passing and failing probes are fit simultaneously
with a signal plus background model. 
The invariant mass shape of the signal is obtained from simulation and the background shape is parameterized using an exponential function. 

Since the pixel detector efficiency is $\abs{\eta}$-dependent, the misidentification ratio is evaluated in bins of $\abs{\eta}$. 
For electrons detected in the EB, three bins are used: 0.0--0.6, 0.6--1.1, 1.1--1.44, with the estimated ratio across the three years ranging from 0.01 in the highest $\abs{\eta}$ region to 0.045 in the lowest $\abs{\eta}$ region. 
In the EE only a single $\abs{\eta}$ range covering 1.57 to 2.5 is used due to the low number of high-\pt electrons detected, and the ratio is estimated to be between 0.0058--0.0165 for the three years. 
Additionally, the efficiencies of certain parts of the pixel detector changed during the three years of data taking; for these regions separate misidentification ratios determined for each year are used. 

The sources of uncertainty for the treatment of this background mainly arise from the limited number of events that fail the probe criteria, which varies across the different $\eta$ regions, and the choice of the signal and background models. 

\subsection{Background due to \texorpdfstring{$\PW(\Pell\PGnell)\PGg$}{W(lnul)gamma}}
\label{sec:wgbkg}

The $\PW(\Pell\PGnell)\PGg$ process contributes to the background when the \PW boson decays leptonically and the charged lepton \Pell is undetected due to acceptance limitations or reconstruction failure. This background is estimated via a simultaneous likelihood fit to data from the SR and to the two control regions, $\text{CR}_\Pe$ and $\text{CR}_\PGm$, which are kinematically similar to the SR but enriched in $\PW(\Pell\PGnell)\PGg$ events by requiring an electron or a muon. These control regions provide a direct constraint on the 
$\PW(\Pell\PGnell)\PGg$ background normalization, reducing reliance on theoretical cross-section predictions. All \ptmiss-related criteria from the SR are transferred to the transverse recoil momentum, $\vec{R}_T$, ensuring phase space consistency. 

Events are not added to the CRs if the lepton is collinear with an energetic jet, so as to exclude events with final-state radiation. Events from 2018 where the lepton is in one of the inoperable HCAL regions are also not included. 

The expected $W(\Pell\PAGnell)\PGg$ contribution is initially modeled using simulation, with the \pt distribution corrected for next-to-next-to-leading order (NNLO) QCD~\cite{kFacQCD_Grazzini} and next-to-leading order (NLO) electroweak~\cite{kFacEWWG_Denner} effects. Free parameters for each \pt bin are used to scale the process yield uniformly across all three regions. 
In the simultaneous fit, the SR-to-CR yield ratios are fixed for each bin, while the overall normalization is set with the data.

\subsection{Other backgrounds}\label{sec:spikes}

Other background sources include the spike signals in the APDs, which can be misidentified as high-$\pt$ photons with a large $\ptmiss$, and bremsstrahlung radiation from beam-halo muons in the ECAL detector that can produce a photon-like signature in both EB and EE.
 
To estimate the number of spike signals within the 3\unit{ns} event window, the distribution of the arrival time of the photon events is fit with signal and background timing distribution templates.  The background template is derived from spikes and beam-halo events. The timing distribution of spikes is derived from a dedicated control region in data, selecting photons with large signals in the central crystals and small signals in the adjacent crystals. The beam-halo template is derived by selecting photons with large MIP$_{\rm{tot}}$ values, and the signal template is obtained from $\PZ\to\Pep\Pem$ events in data. 

The estimated number of spikes obtained from this fit ranges from 9.4 to 20.8 across the three years of data-taking, with an uncertainty no more than $17\%$, while the contribution from beam-halo to the total estimated background is negligible. 
A more detailed description of the spike background estimation procedure can be found in Ref.~\cite{cms_dm_search_2019}. 

In the EE, the background due to bremsstrahlung radiation by beam-halo muons is the predominant source of background and the strategy used in the EB to remove this background is ineffective in the EE due to its geometry. In this region, beam-halo muons cross the EE nearly parallel to the beamline and the bremsstrahlung photons that they produce deposit energy in a small number of adjacent crystals making them difficult to distinguish from photons produced at the interaction point.
To suppress this background the score of a dedicated BDT using 
the angle between a line joining the hits in the EE, and either the HCAL endcap or the ES, and the beam line, and the energy deposited in both the HE and ES, as well as the
ECAL shower shape variables are used to separate signal from beam-halo events.
The application of this BDT reduces the beam-halo background to $<$0.2\% of its initial contribution in all the 3 years.

The remaining contribution of the beam halo to the total background is evaluated by fitting the $\phi$ distribution with the signal template, obtained from simulation, and
a background template, extracted from data by inverting the BDT tagger requirement. 
 
\section{Systematic uncertainties} 
\label{sec:unc}
For simulated signal and background sources, uncertainties are assigned to the estimated luminosity of the dataset, the efficiencies of the trigger and photon identification, the calibration of the jet energy scale and resolution, the calibration of the photon energy scale and resolution, and the modeling of the pileup interactions. In addition, uncertainties are assigned for the choice of PDF set and for the missing higher orders in QCD calculations by varying of the renormalization and factorization scales, $\mu_{\mathrm{R}}$ and $\mu_{\mathrm{F}}$. These uncertainties can affect the shape of the $\ptgamma$ spectrum and the event yield. The large uncertainties in the background due to electrons misidentified as photons, and jets misidentified as photons are due to the limited number of events in the control regions where the misidentification ratio is measured. 

The uncertainties in the integrated luminosity are estimated to be 1.3\% (2016), 2.2\% (2017), and 2.5\% (2018)~\cite{lumi2016,lumi2017, lumi2018}. In addition, the total integrated luminosity has an uncertainty of 1.6\% reflecting the uncorrelated time evolution of some systematic effects. Simulated background events are weighted in order to match the expected pileup distribution obtained using the inelastic $\Pp\Pp$ interaction cross section, $69.2\pm4.6$\unit{mb}~\cite{CMS:2018mlc}.

The uncertainties in the correction factors that account for the difference in the lepton ID efficiencies between data and simulated data are below 0.01\%. An uncertainty of between 3.5\% and 4.7\% is assigned to the correction factors, to account for the difference between the photon ID efficiencies in data and simulated data.  

To assess the impact of the uncertainties in the photon energy scale and resolution the electron and photon energies are varied by around 0.1 (0.3)\% in the EB (EE)~\cite{CMS:2020uim}. 
As the signal photon is required to not have a pixel detector seed in the tracker, the uncertainty due to mismodeling of the efficiency of this requirement is obtained in the data and is under 0.9\%.
 
{\tolerance=1200
Jet $p_T$ measurement uncertainties are propagated by varying the energy corrections within their estimated uncertainties. 
This affects the yields due to the phase space selection on $\Delta \phi_\text{min}(\ptvec^\text{jet}, \ptvecmiss)$ in the SR and $\Delta \phi _{\min}(\ptvec^\text{jet}, \vec{R}_\mathrm{T})$ in the CRs, where only jets with \pt larger than 30\GeV are considered. The jet energy variations and the corresponding \ptvecmiss variations are correlated and are thus found simultaneously in the calculations of the yields. 
The L1 ECAL trigger inefficiency due to timing misalignment~\cite{CMS:2020cmk} is not included in the simulation; weights to correct for this are extracted from data. The uncertainty associated with these weights is about 1.3\% for 2016 and 1.9\% for 2017 and 2018.  
The uncertainty in the $\Pe\to\PGg$ misidentification ratio and the jet $\to\PGg$ misidentification, described in Sections~\ref{sec:lep_fakes} and~\ref{sec:jet_fakes}, ranges from 4\% to 20\%. 
\par}

Uncertainties in the NNLO QCD corrections for $\PZ(\PGn\PAGn)\PGg$ production are assigned for the $\mu_\mathrm{F}$ and $\mu_\mathrm{R}$ scales calculated with \textsc{MATRIX}~\cite{matrix_Grazzini2018}, and for shape variations following the prescription in Ref.~\cite{nnloQcdUncLindert2017}. 
Uncertainties in NLO electroweak corrections are assigned as prescribed in Refs.~\cite{kFacEWZG_Denner, kFacEWWG_Denner}. 
The simulated background estimates use the NNPDF 3.1 Hessian PDF set. The uncertainties due to PDF and \alpS are estimated following the PDF4LHC recipe~\cite{butterworth_pdf4lhc_2016}. 
The uncertainties due to missing higher QCD orders, which includes the renormalization and factorization scale choice, and QCD NNLO corrections, are estimated by independently varying the two scales by factors of 0.5 and 2 with the constraint $0.5 \leq \mu_\mathrm{R}/\mu_\mathrm{F} \leq2$, and taking the envelope (max and min) of the 7-point variations. 
The values of these systematic uncertainties are given in Table~\ref{tab:sys_unc}. The uncertainties of the photon ID corrections and the jet energy scale are the two leading sources of systematic uncertainty in this analysis.

\begin{table}[ht]
\centering
\topcaption{Sources of systematic uncertainties and their estimated contributions. The $\checkmark{}$ symbol indicates full correlation between the data-taking years.} 

\label{tab:sys_unc}
\begin{tabular}{lcc}
\hline
\multicolumn{1}{l}{Source}                              &  & \multicolumn{1}{c}{Correlated between years}                 \\ \hline
\multicolumn{3}{c}{Experimental}                                                                                                                                                      \\[\cmsTabSkip]
\multicolumn{1}{l}{Integrated luminosity}                          & \multicolumn{1}{c}{1.3--2.5\%}        & \multicolumn{1}{c}{Partially}           \\
\multicolumn{1}{l}{Pileup}                             & \multicolumn{1}{c}{0.4--2.2\%}    & \multicolumn{1}{c}{\checkmark{}}  \\
\multicolumn{1}{l}{Lepton ID corrections}                       & \multicolumn{1}{c}{$<0$.01\%}                  & \multicolumn{1}{c}{\checkmark{}} \\
\multicolumn{1}{l}{Photon ID corrections}                       & \multicolumn{1}{c}{3.5--4.7\%}    & \multicolumn{1}{c}{\checkmark{}}  \\
\multicolumn{1}{l}{Photon energy scale and resolution}  & \multicolumn{1}{c}{0.2--1.1 \%}   & \multicolumn{1}{c}{\checkmark{}}  \\
\multicolumn{1}{l}{Photon pixel veto corrections}               & \multicolumn{1}{c}{0.1--0.9\%}    & \multicolumn{1}{c}{\checkmark{}}  \\
\multicolumn{1}{l}{Jet energy scale and resolution}     & \multicolumn{1}{c}{0.2--2.8\%}      & \multicolumn{1}{c}{Partially}   \\
\multicolumn{1}{l}{L1 ECAL inefficiency}                   & \multicolumn{1}{c}{0.1--1.9\%}     & \multicolumn{1}{c}{\checkmark{}}  \\
\multicolumn{1}{l}{Jet$\to\PGg$ misid. ratio}   & \multicolumn{1}{c}{2--20\%}          & \multicolumn{1}{c}{\checkmark{}}  \\
\multicolumn{1}{l}{$\Pe\to\PGg$ misid. ratio}     & \multicolumn{1}{c}{13--45\%}         & \multicolumn{1}{c}{\checkmark{}}  \\[\cmsTabSkip]
\multicolumn{3}{c}{Theoretical}                                                                                                                                         \\[\cmsTabSkip]
\multicolumn{1}{l}{NLO EW corrections}         & \multicolumn{1}{c}{2.7\%}             & \multicolumn{1}{c}{\checkmark{}}  \\
\multicolumn{1}{l}{PDF and \alpS}                & \multicolumn{1}{c}{1.5\--1.7\%}      & \multicolumn{1}{c}{\checkmark{}}  \\
\multicolumn{1}{l}{Renormalization and factorization scale} & \multicolumn{1}{c}{6--7.2\%}        & \multicolumn{1}{c}{\checkmark{}}  \\ \hline
\end{tabular}
\end{table}

\section{Statistical analysis}
\label{sec:statAn}

Cross section measurements and the constraints on the anomalous coupling parameters are evaluated by constructing
a binned profile likelihood function to determine the best fit parameters to the data using the COMBINE tool of CMS~\cite{CMS:2024onh}. 
The binned likelihood function is defined as:
\begin{equation}
\begin{aligned}
    \mathcal{L}(\vec{d}\mid \vec{\mu},\, \vec{r}^{\PW\Pgg},\, \vec{\theta})~~=&~~~~\prod_{i} \biggl[ \mathcal{P}\big(d_{i} | s_{i}(\vec{\mu}, \vec{\theta}) + r^{\PW\Pgg}_{i} \, N^{\text{SR},\, \PW\Pgg}_{i}(\vec{\theta}) + b_{i}(\vec{\theta}) \big) \\
    \times ~&\prod_{K=\Pe,\Pgm} \mathcal{P}\big( d^{K}_{i} | r^{\PW\Pgg}_{i} \,N^{K,\,\PW\Pgg}_{i}(\vec{\theta}) + b^{K}_{i}(\vec{\theta}) \big) \biggr] \\
    \times ~&~~~~\prod_{j} \mathcal{N}(\theta_j), 
    \label{eqn:finalNLL}
\end{aligned}
\end{equation}
where $i$ denotes the index for the $\pt^{\PGg}$-bins, $\vec{\mu}$ represents the parameters of interest, which are either the aNTGC coupling parameters ($h_3$ and $h_4$) or the cross sections, while $\vec{\theta}$ denotes the nuisance parameters associated with systematic uncertainties. The uncertainties due to limited statistics of the simulated samples are incorporated using Barlow-Beeston Lite method~\cite{Barlow:1993dm}. 
The symbol $\mathcal{P}$ indicates the Poisson distribution and $\mathcal{N}$ the normal distribution. Within bin $i$, the observed event counts in the SR and $\text{CR}_K$ (indexed by $K=\Pe, \PGm$) are denoted by $d_i$ and $d_i^K$, respectively. The contributions from the $\PW(\Pell\PGnell)\PGg$ process, estimated by simulation, are $N^{\text{SR}, \PW\Pgg}_i$ and $N^{K, \PW\Pgg}_i$, while the contributions from all other background processes are $b_i$ and $b_i^K$ in the SR and $\text{CR}_K$, respectively. A free parameter, $r^\textit{\PW\Pgg}_i$, scales the simulated estimates of the $\PW(\Pell\PGnell)\PGg$ process within each bin $i$ in all three regions, thereby constraining the rate of this background. The signal terms $s_i$ are defined in terms of the parameters of interest, which are extracted by a simultaneous maximum likelihood fit to the data. Table~\ref{tab:yields_full} shows the post-fit event yields in the photon-$\pt$ spectrum for the combined phase space extracted after a simultaneous fit to the data in the signal and control regions. As expected, due to the high-$\pt$ of the photon, the EB has significantly more events than the EE. Separately, as a confirmation, in a fit to only the control regions the estimated number of $\PW(\Pell\PGnell)\PGg$ events in the signal region is $321 \pm 19$ in the barrel region and $69.8 \pm 6.3$ in the endcap region, which is consistent with the yields obtained in the simultaneous fit.

\begin{table}
\centering
\topcaption{Post-fit event yields in the photon-$\pt$ spectrum for the combined phase space. These numbers are extracted after a simultaneous fit to the data in the signal and control regions. The minor background include contributions from $\text{nonfiducial}$ $\PZ(\PGn\PAGn)\PGg$, $\PZ/\PGg^{\ast}(\Pell^+\Pell^-)\PGg$, $\ttbar\PGg$, $\PW\PV\PGg$, single-$\PQt$+$\PGg$, $\PV\PV$, and single-$\PW$.}
\label{tab:yields_full}

\begin{tabular}{lcc} 
\hline
Predicted & Barrel& Endcaps\\ 
\hline
Fiducial $\PZ\PGg$ & $1456 \pm 73$ & $305 \pm 17$ \\ 
Non-fiducial~$\PZ\PGg$  & $72.2 \pm 9.8$   & $21.8 \pm 2.9$   \\
$\Pe\to\PGg$ & $404 \pm 53$  & $37.4 \pm 11.1 $  \\ 

$\PW(\Pell\PGnell)\PGg$& $319 \pm 29$& $69.8 \pm 8.1$ \\ 

jet$~\to\PGg$ & $34.5 \pm 0.8$ & $12.2 \pm 0.6$  \\ 

Minor backgrounds                             & $153 \pm 28$                         & $81 \pm 89$                          \\ 

Spikes                                       & $40.3 \pm 3.7$                       & \NA                                     \\ 

Beam halo                                    & \NA                                   & $9.5 \pm 0.5$                        \\

Total Predicted                                    & $2480 \pm 51$                       & $539 \pm 184$                          \\ 
Observed                                     & $2453$                             & $568$                                \\ 
\hline
\end{tabular}

\end{table}

\section{Results}
\label{sec:res}
\subsection{Fiducial cross section and differential cross section}
\label{sec:fid_xsec}
The inclusive differential cross section for the $\PZ(\PGn\PAGn)\PGg$ process is measured with a likelihood-based unfolding technique~\cite{unfolding_GAPONENKO2020163612} that accounts for the impact of the detector acceptance, efficiency, and resolution. 
The measurements correspond to a fiducial region at the generator level of the particles before any interaction with the detector.  This region is defined by criteria that closely match the detector level phase space of the SR. The selections applied on the generator level variables of the simulated signal are: 
\begin{itemize}
\item $(\abs{\eta^{\PGg}}<1.44) \cup  (1.57 < \abs{\eta^{\PGg}} < 2.5)$, 
\item $p_T^{\PGg}>225\GeV$, 
\item $\ptmiss>200\GeV$, 
\item $\frac{p_T^{\PGg}}{\ptmiss}<1.4$,
\item $\Delta \phi (\ptvecmiss, \PGg)>2$, 
\end{itemize}

where $\ptvecmiss$ is the magnitude of the vector sum $\vec{p}_\mathrm{T}$  of the two neutrinos coming from the 
$\PZ$ boson at the generator level, and $\Delta \phi (\ptvecmiss, \PGg)$ refers to the azimuthal angle between the transverse momentum vector of the generator-level photon and $\ptvecmiss$. 

We construct the response matrix, $R_{ij}$, which models the detector response to the events produced in the fiducial region to quantify the probability with which an event produced with a generator level photon \pt in bin $\textit{j}$ will be observed with detector-level photon \pt in bin $i$. This is used to account for bin migration due to the detector energy resolution, efficiency and acceptance. 
We calculate the number of events from the signal simulation, $N_{ij}^\text{detector}$, in the detector-level bin $i$ that are actually produced in the generator level bin $j$. The sample also provides the total number of generator level fiducial events, $N_j^\text{generator}$, produced in the generator level bin $j$. The response matrix is then defined as:
\begin{equation}
R_{ij} = \frac{N_{ij}^{\text{detector}}}{N_j^\text{generator}}.
\label{eq:response_matrix_def}
\end{equation}
This response is used to model the signal, $s_i$, within the likelihood function as given in Section~\ref{sec:statAn}. The number of expected signal events in detector bin $i$ is expressed as a sum of contributions from the generator level bins ${ \textit{j} }$, each associated with unknown cross sections ${ \sigma_\textit{j} }$. Additionally, there is a minor contribution, $s_i^\text{nonfiducial}$, originating from signal process events occurring outside of the fiducial region. These nonfiducial events are treated as background in this measurement. The full signal model is:
\begin{equation}
s_i = s_i^\text{nonfiducial} + \sum_{j} \mathcal{L} \, \sigma_j \, R_{ij},
\label{eq:signal_diff_xsec_response}
\end{equation}
where $\mathcal{L}$ is the integrated luminosity. The value of $\mathcal{L} \, \sigma_\textit{j}$ is the true number of events produced at the generator level in bin $\textit{j}$, which is reduced by the detector response $R_{ij}$ in the sum. The terms $\{ \sigma_\textit{j} \}$ are treated as free parameters and obtained from a maximum likelihood fit to derive the differential cross sections. The chosen bin boundaries for the photon's generator level \pt are:
225, 275, 350, 450, 600, 800, and 1500 $\GeVns$. 

\begin{figure}[ht]
    \centering
    \includegraphics[width=0.49\linewidth]{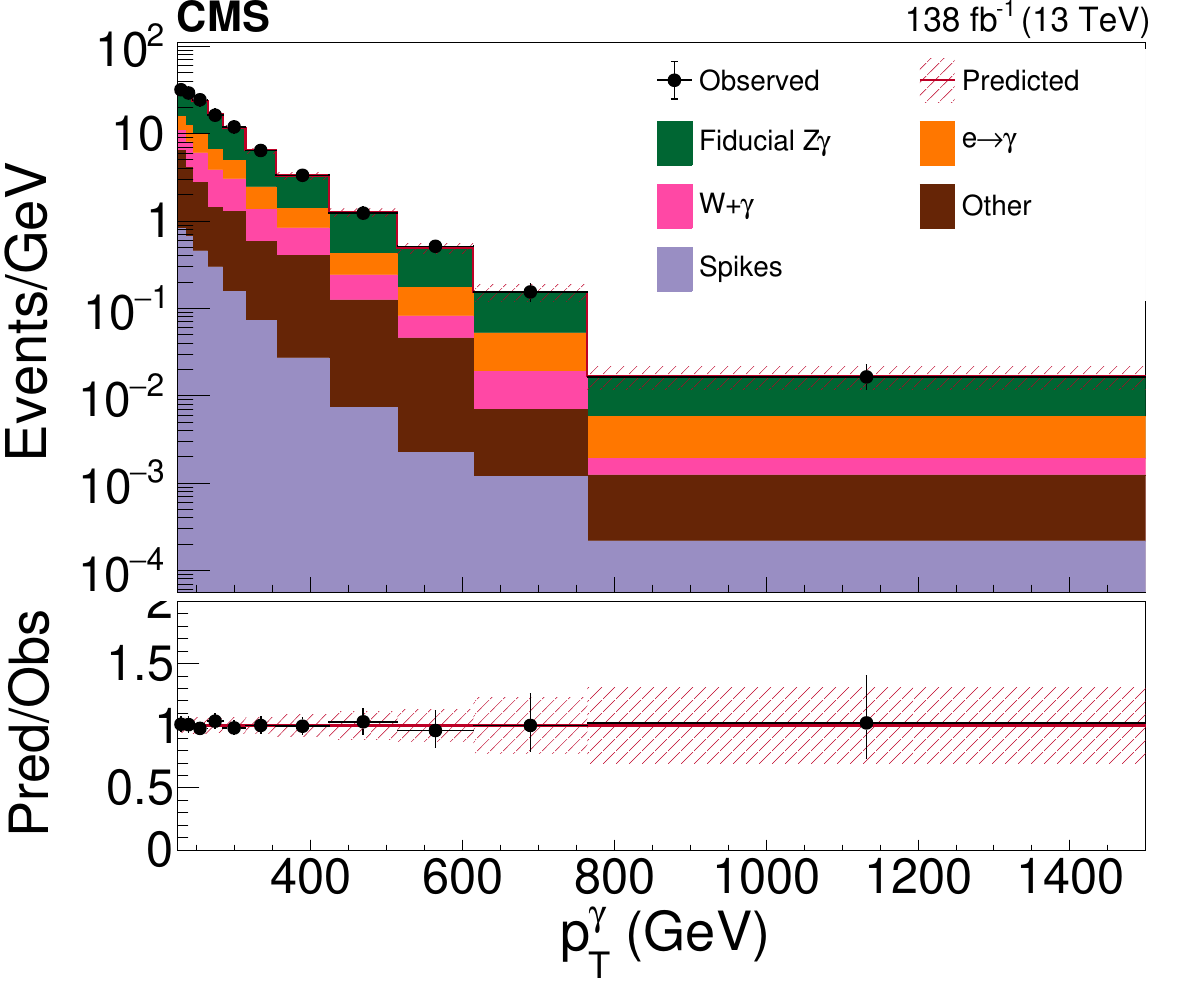}
    \includegraphics[width=0.49\linewidth]{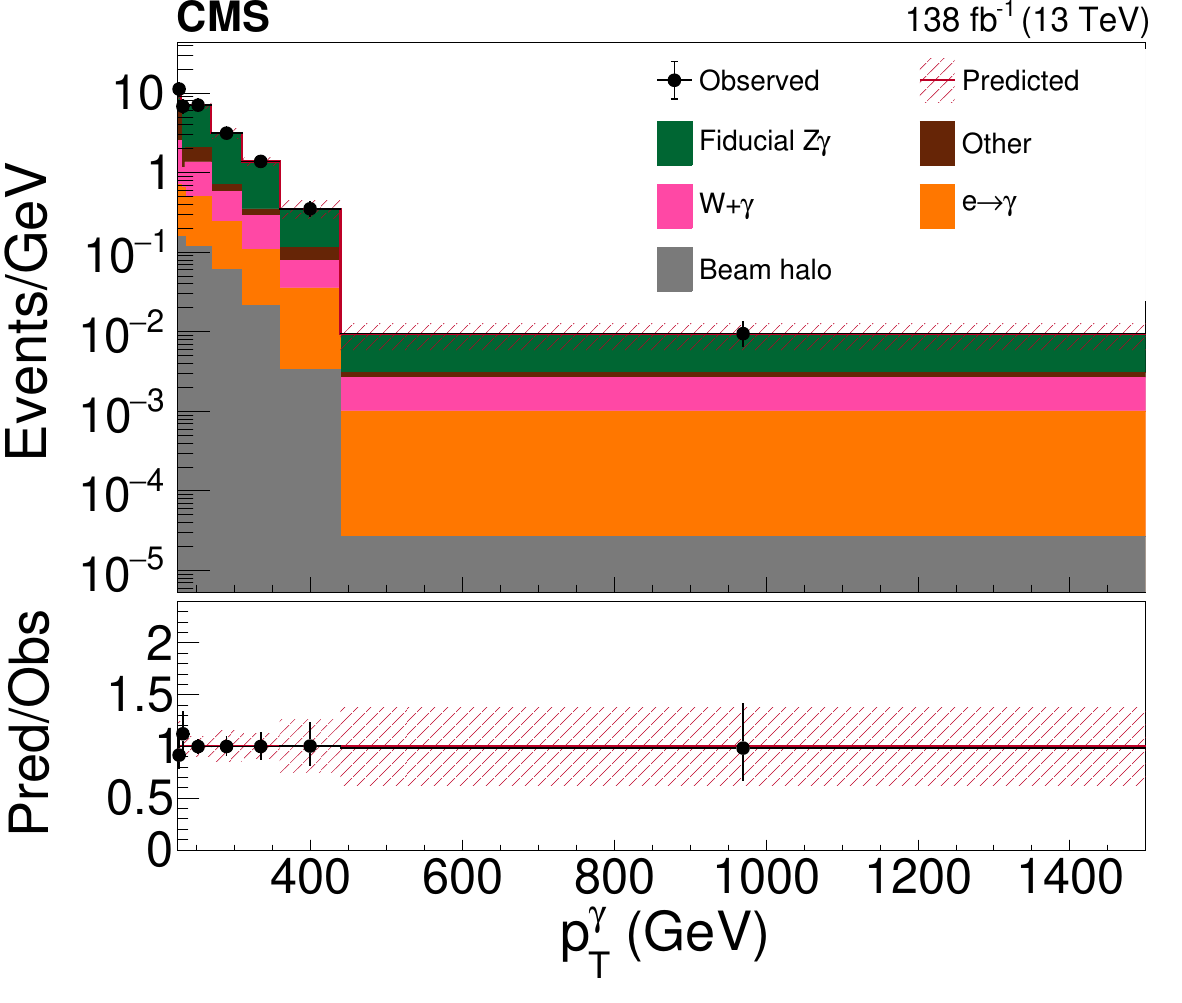}
    \caption{The post-fit reconstruction-level distributions in the signal region after the combined (barrel+endcap) differential cross section fit (left: barrel region, right: endcap region). The Predicted distribution is the full model with all parameters fixed to their post-fit values. Note in particular that the free cross section parameters for the signal are also fixed to their post-fit values and then propagated through the response matrix. The bars on the black data points indicate the statistical uncertainty in the data, while the red band shows the total post-fit model uncertainty.}
    \label{fig:dist_pt_diff_xsec} 
\end{figure}

The measured cross-sections and the predictions at NLO and NNLO in QCD using \MGvATNLO~\cite{MadGraph} and \textsc{MATRIX}~\cite{matrix_Grazzini2018} are shown in Table~\ref{tab:fid_xsec}. 
Due to nuisance parameters being pulled to different values in different fit configurations, the sum of the cross section in the EB and EE is not exactly equal to the combined cross section. 
Figure~\ref{fig:dist_pt_diff_xsec} shows the observed and fitted distributions of photon \pt in the barrel and endcap regions of the phase space obtained separately after performing the fit for the measurement of the differential cross sections for the combined phase space. For these distributions, the $\PW(\ell\nu_{\ell})\PGg$ background is estimated only from the control region fit. 

\begin{figure}[ht]
    \centering
    \includegraphics[width=0.6\linewidth]{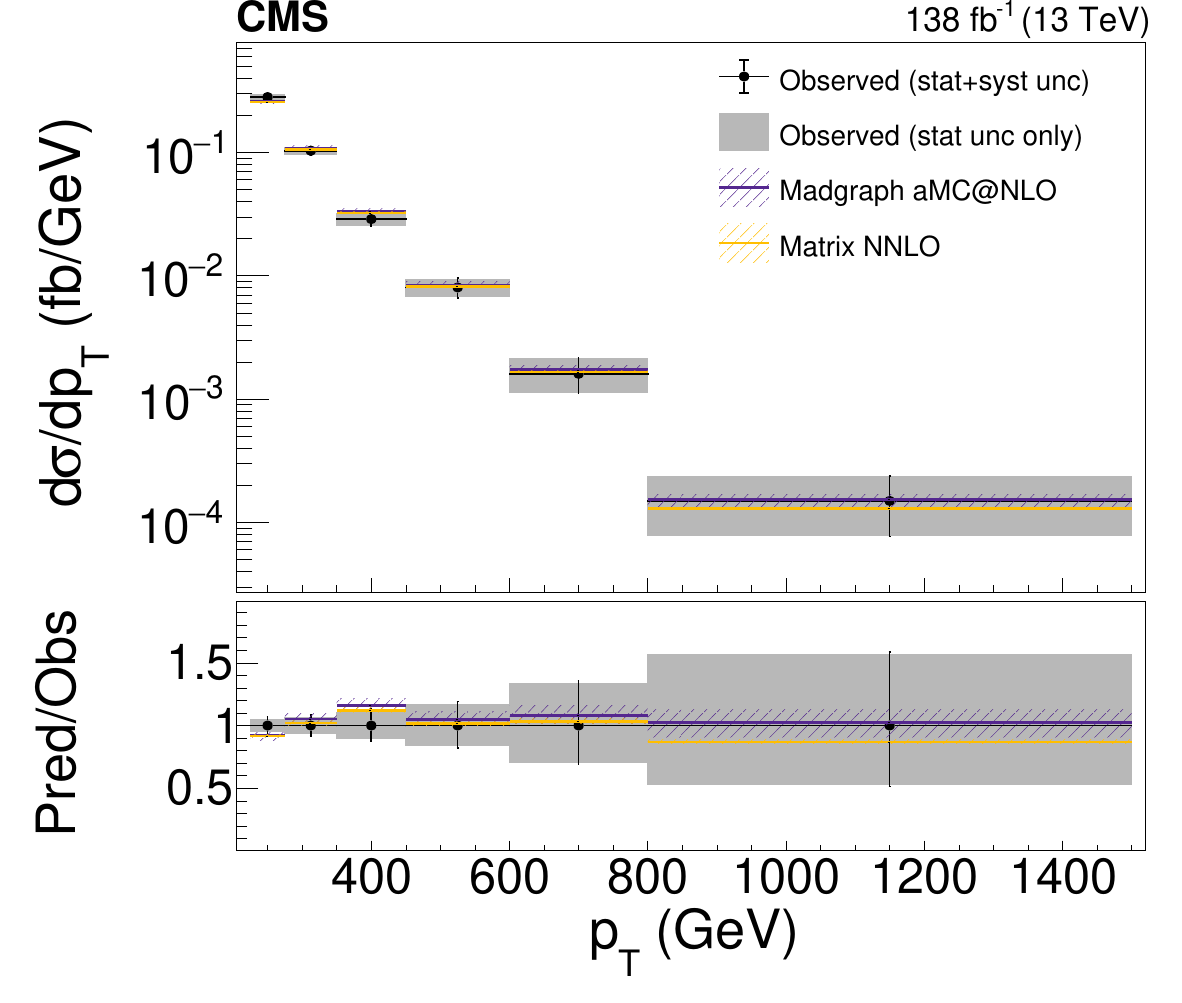}
\caption{Measured and predicted differential cross sections as function of $\pt^{\PGg}$ for the combined fiducial phase space. The colored bands associated with the predictions indicate the theoretical uncertainties associated with the  \MGvATNLO and \textsc{MATRIX} generators.
For the measured cross sections, the grey band indicates the statistical uncertainty of the fit, while the error bars on the black data points indicate the total uncertainty of the fit, including both statistical and systematic components.}
\label{fig:diff_xsec_pT}
\end{figure}

{\tolerance=1200
The measured differential cross sections and the theoretical predictions of \MGvATNLO and \textsc{MATRIX} as a function of $\pt^{\PGg}$ are shown in Fig.~\ref{fig:diff_xsec_pT} and are listed in Table~\ref{tab:diff_xsec_list}. The theoretical predictions have uncertainties from renormalization and factorization scale variation, as well as numerical precision. 
Additionally, \MGvATNLO includes the uncertainties of parton density functions.  
While the theoretical predictions are all higher than the measured cross-sections, the differences are all within the uncertainties, except for the \pt bins 275-350\GeV and 350-450\GeV, where the differences are $\approx$2$\sigma$, with the NNLO(\textsc{MATRIX}) predictions in better agreement than those from \MGvATNLO. 
The low \pt bins have the highest correlation where it ranges between 0.20 and 0.27. 
The binning in detector-level photon \pt is optimized using the Bayesian Block algorithm~\cite{Pollack:2017srh}. 
\par}

\begin{table}[ht]
\centering
\topcaption{Measured and predicted cross sections in femtobarns evaluated in the barrel, the endcaps, and the combined fiducial phase space. The predicted cross sections are calculated at NLO and NNLO in QCD. Due to nuisance parameters being pulled to different values in different fit configurations, the sum of the cross section in the EB and EE is not exactly equal to the combined cross section.}
\label{tab:fid_xsec}
\begin{tabular}{cccc}
\hline
                    Region                                                                              & \multicolumn{1}{c}{Measured}                 & \multicolumn{1}{c}{NLO (\MGvATNLO)} & NNLO (\textsc{MATRIX}) \\ \hline
	\multicolumn{1}{c}{\begin{tabular}[c]{@{}c@{}}Barrel\\ $\abs{\eta}~<~1.4442$\end{tabular}}          & \multicolumn{1}{c}{$16.7 ^{+1.0}_{-1.0}$} & \multicolumn{1}{c}{$19.6 ^{+0.7}_{-0.7}$}           & $19.3 ^{+0.3}_{-0.3} $         \\ 
		\multicolumn{1}{c}{\begin{tabular}[c]{@{}c@{}}Endcaps\\ $1.566~<~\abs{\eta}~<~2.5$\end{tabular}}     & \multicolumn{1}{c}{$7.8 ^{+0.8}_{-0.7}$}  & \multicolumn{1}{c}{$6.4 ^{+0.3}_{-0.3}$}            & $6.21 ^{+0.07}_{-0.09}$ \\ 
			\multicolumn{1}{c}{\begin{tabular}[c]{@{}c@{}}Total\end{tabular}} & \multicolumn{1}{c}{$23.3 ^{+1.4}_{-1.3}$ } & \multicolumn{1}{c}{$26.1 ^{+1.0}_{-1.0}$ }           & $25.4 ^{+0.4}_{-0.3}$         \\ \hline
\end{tabular}
\end{table}

\begin{table}
\centering
\renewcommand{\arraystretch}{1.3}
\topcaption{Measured and predicted cross sections in femtobarns in bins of $\pt^{\PGg}$.}
\label{tab:diff_xsec_list}
\begin{tabular}{cccc}
\hline
$\pt$-range (\GeVns) & Measured & NLO (\MGvATNLO)  & NNLO (\textsc{MATRIX})\\
\hline
225--275 & ${12.45}^{+0.91}_{-0.87}$&${12.88}^{+0.42}_{-0.49}$ & ${12.83}^{+0.17}_{-0.15}$\\

275--350 & ${6.95}^{+0.60}_{-0.57}$ & ${8.14}^{+0.31}_{-0.30}$ & ${7.89}^{+0.16}_{-0.15}$\\

350--450 & ${2.61}^{+0.35}_{-0.33}$ & ${3.34}^{+0.17}_{-0.11}$ & ${3.22}^{+0.07}_{-0.06}$\\

450--600 & ${1.08}^{+0.21}_{-0.20}$ & ${1.26}^{+0.08}_{-0.05}$ & ${1.22}^{+0.02}_{-0.02}$\\

600--800 & ${0.282}^{+0.102}_{-0.089}$ & ${0.345}^{+0.026}_{-0.019}$ & ${0.331}^{+0.007}_{-0.005}$\\

800--1500 & ${0.092}^{+0.055}_{-0.046}$ & ${0.107}^{+0.011}_{-0.012}$ & ${0.0909}^{+0.0020}_{-0.0014}$\\
\hline
\end{tabular}
\end{table}

\subsection{Limits on anomalous couplings}

\begin{figure}[ht]
    \centering
    \includegraphics[width=0.49\linewidth]{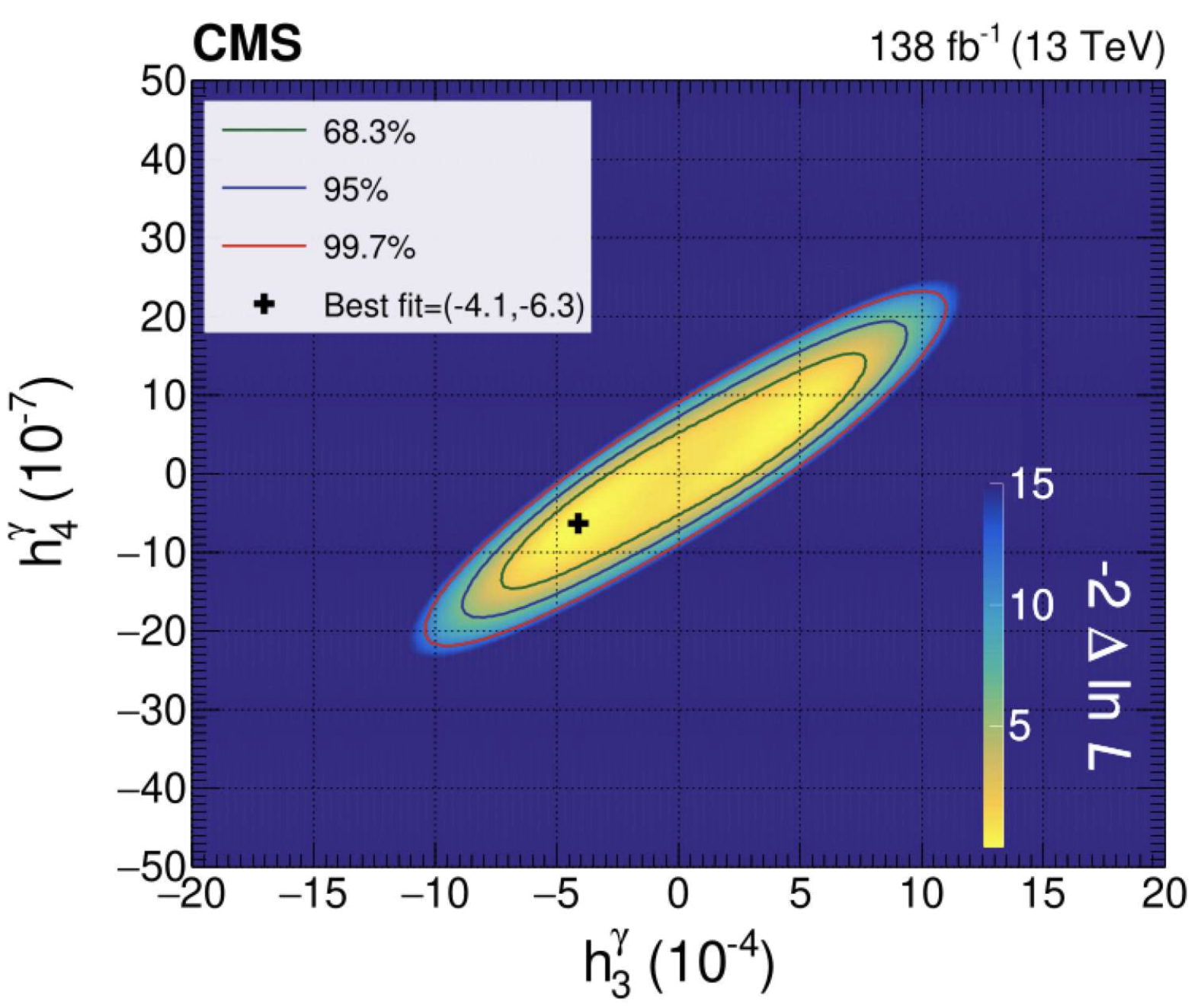}
    \includegraphics[width=0.49\linewidth]{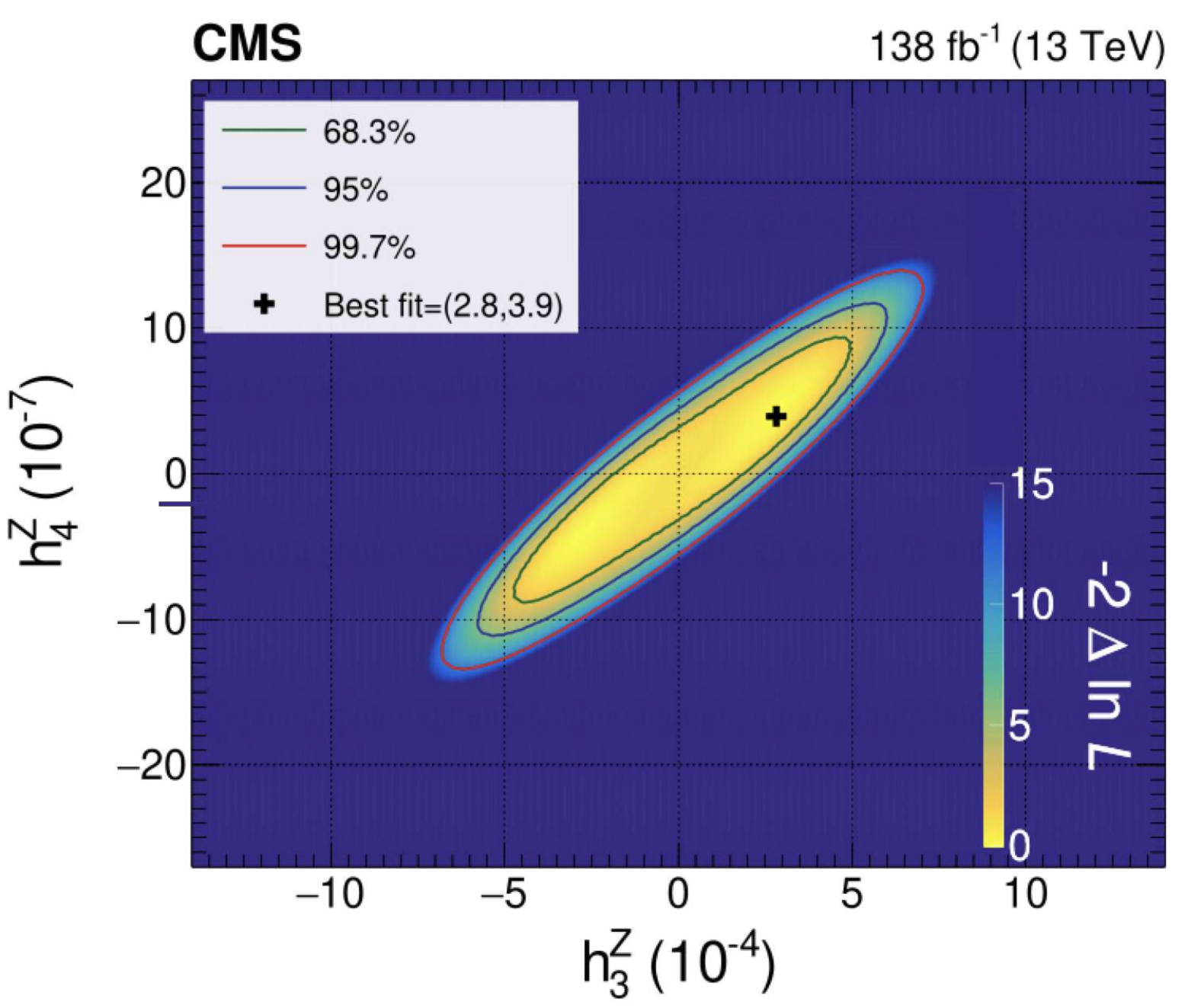}
    \caption{Confidence contours of the aNTGC parameters $h_3^\PGg, h_4^\PGg$ (left) and $h_3^\PZ, h_4^\PZ$ (right) with the best fit points, which are the local minima of the negative log likelihood.} 
    \label{fig:antgc_contours}
\end{figure}

\begin{figure}[ht]
    \centering
    \includegraphics[width=0.49\linewidth]{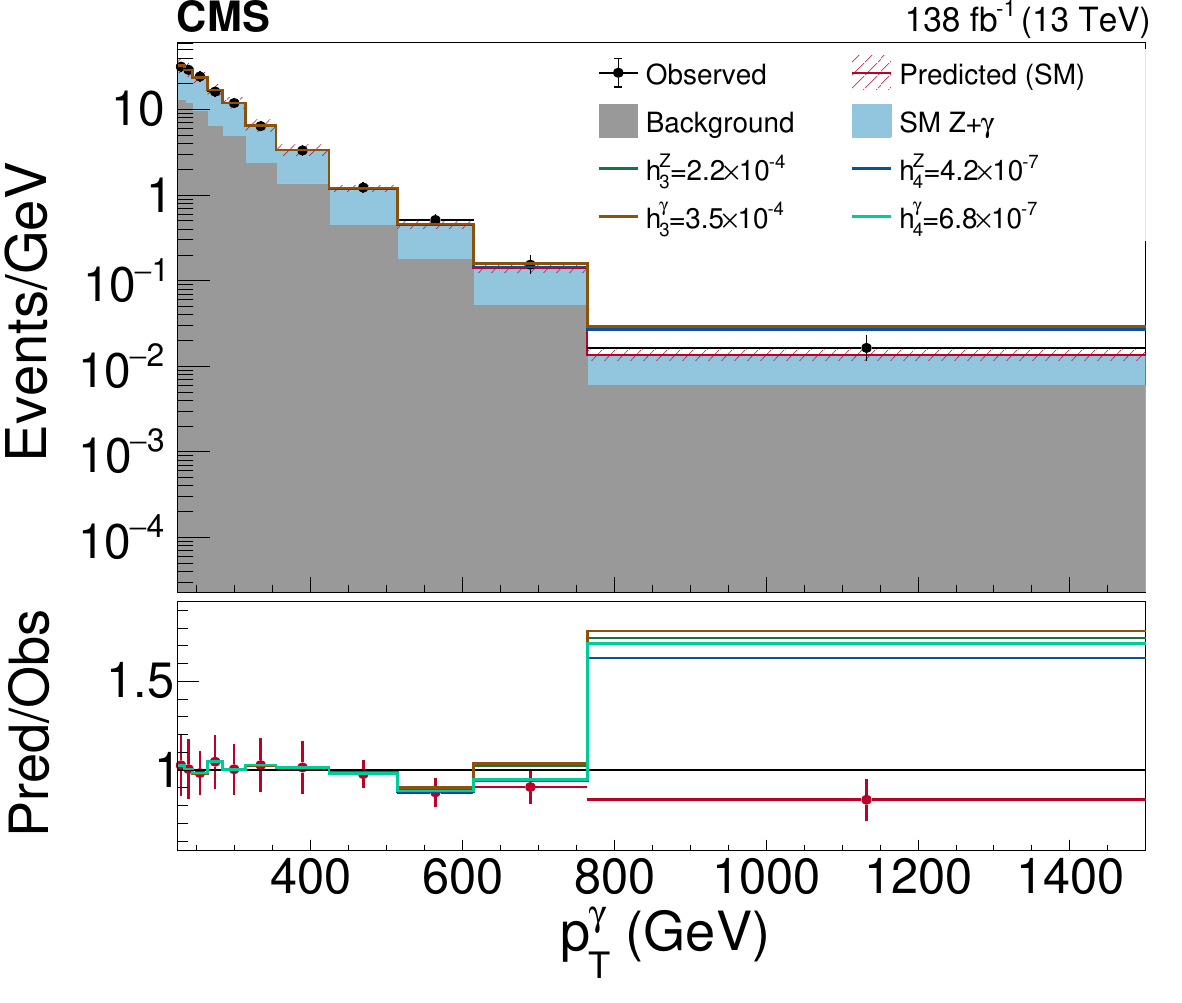}
    \includegraphics[width=0.49\linewidth]{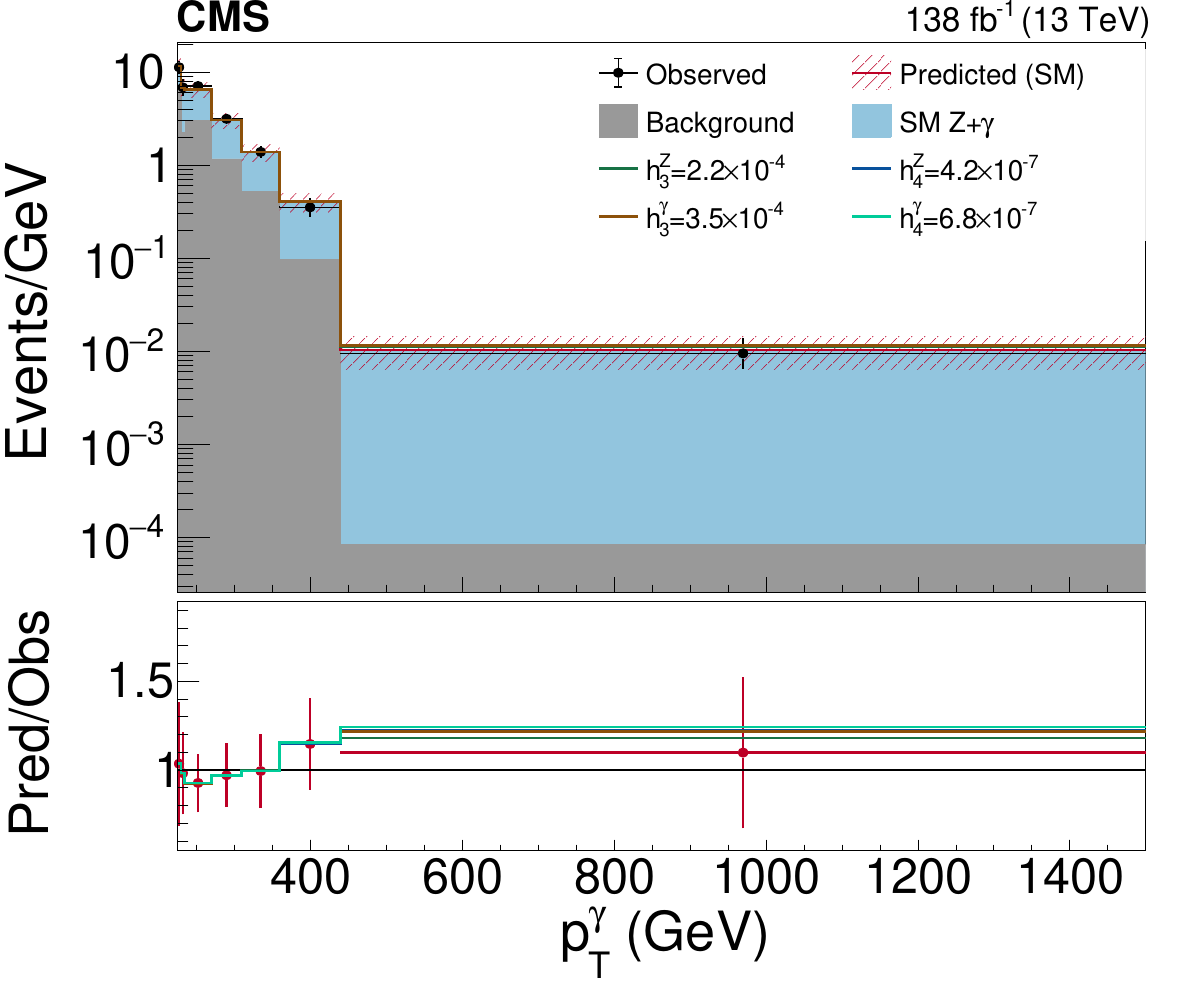}
    \caption{Comparison of the predicted distribution of $\pt^{\PGg}$ under the SM-only hypothesis in the barrel (left), and in the endcaps (right) phase space with the predictions at the upper boundaries of the observed 1D confidence intervals of the coupling parameters. The lower panels compare the SM (shown in red) and the aNTGC predictions, with the observed $\pt^{\PGg}$ distribution (shown in black). The uncertainties on the data are statistical only; the uncertainties on SM predictions include renormalization and factorization scale variations.
}
    \label{fig:antgc_pt_comparison}
\end{figure}

We parameterize the aNTGC prediction for the SR by fitting the simulated yields to 
different pairs of coupling values with a two-dimensional quadratic function of $h_{3}^{\PZ}$ and $h_{4}^{\PZ}$ as a function of $\pt^{\PGg}$. This parameterization is then used to fit the likelihood simultaneously in the EB and the EE. To calculate the likelihoods, the binning chosen is the same as that used for the histograms in Fig.~\ref{fig:dist_pt_diff_xsec}. 
The coefficients of the parameterization are varied to determine the systematic uncertainties. The confidence contours obtained are shown in Fig.~\ref{fig:antgc_contours}. 
As the interference term in the Lagrangian becomes negligible at high transverse momentum, the measurement is insensitive to the sign of the couplings, we therefore quote the best fit points in the data at the local minima. 
In addition, one-dimensional confidence intervals for each parameter, with the other parameters fixed at zero, are shown in Table~\ref{tab:oneD_limits}. In Fig.~\ref{fig:antgc_pt_comparison}, the predicted $\pt^{\PGg}$ distribution after performing the fit under the SM-only hypothesis is compared with the predictions at the upper boundaries of the observed 1D confidence intervals of the coupling parameters.

\begin{table}
\centering
\topcaption{Expected and observed 95\% confidence level intervals of the parameters of anomalous couplings (with the other parameters fixed at zero). }
\label{tab:oneD_limits}
\begin{tabular}{lll}
\hline
Parameter                               & Expected    & Observed                         \\ 
\hline
$h_3^\PGg \times 10^{4}$  & ($-$2.8, 2.9) & ($-$3.4, 3.5)                      \\

$h_4^\PGg \times 10^{7}$ & ($-$5.9, 6.0) & ($-$6.8, 6.8)                      \\

$h_3^{Z} \times 10^{4}$      & ($-$1.8,1.9)  & ($-$2.2, 2.2)                      \\

$h_4^\PZ \times 10^{7}$      & ($-$3.7, 3.7) & ($-$4.1, 4.2)         \\            
\hline
\end{tabular}
\end{table}

\section{Summary}
\label{sec:summary}
We have measured the inclusive and differential cross sections of $\PZ(\PGn\PAGn)\PGg$ in proton-proton collisions at $\sqrt{s}=13\TeV$ using data collected with the CMS detector between 2016 and 2018, corresponding to an integrated luminosity of 138\fbinv. The fiducial $\PZ\PGg$ cross section, where a photon with a \pt greater than 225\GeV is produced in association with a \PZ, and the \PZ decays to a $\Pgn\Pagn$ pair ($\PZ(\Pgn\Pagn)\PGg$), is measured to be $23.3^{+1.4}_{-1.3}\unit{fb}$, in agreement, within uncertainties, with the standard model prediction, which are $26.07^{+0.96}_{-0.97}\unit{fb}$ from \MGvATNLO and $25.45^{+0.41}_{-0.33}\unit{fb}$ from \textsc{MATRIX} (NNLO). The differential $\PZ(\PGn\PAGn)\PGg$ cross section as function of photon transverse momentum has been measured and compared with the standard model predictions computed at next-to-leading and next-to-next-to-leading orders in perturbative quantum chromodynamics. With these data we have probed the presence of anomalous couplings that affect the $\PZ\PZ\PGg$ vertex by using the \pt spectrum of the photons as the sensitive variable. No evidence is found for anomalous neutral trilinear gauge couplings in $\PZ\PGg$ production. The observed 95\% confidence level intervals for $h_3^{\PGg}$ and $h_4^{\PGg}$ are determined to be ($-$3.4, 3.5)$\times10^{-4}$ and ($-$6.8, 6.8)$\times10^{-7}$; and for $h_3^{\PZ}$ and $h_4^{\PZ}$ ($-$2.2, 2.2)$\times10^{-4}$ and ($-$4.1, 4.2)$\times10^{-7}$, respectively. These are stricter than the earlier limits on $h_3^{\PGg}$, $h_3^{\PZ}$, and $h_4^{\PZ}$.  

\begin{acknowledgments}
We congratulate our colleagues in the CERN accelerator departments for the excellent performance of the LHC and thank the technical and administrative staffs at CERN and at other CMS institutes for their contributions to the success of the CMS effort. In addition, we gratefully acknowledge the computing centers and personnel of the Worldwide LHC Computing Grid and other centers for delivering so effectively the computing infrastructure essential to our analyses. Finally, we acknowledge the enduring support for the construction and operation of the LHC, the CMS detector, and the supporting computing infrastructure provided by the following funding agencies: SC (Armenia), BMBWF and FWF (Austria); FNRS and FWO (Belgium); CNPq, CAPES, FAPERJ, FAPERGS, and FAPESP (Brazil); MES and BNSF (Bulgaria); CERN; CAS, MoST, and NSFC (China); MINCIENCIAS (Colombia); MSES and CSF (Croatia); RIF (Cyprus); SENESCYT (Ecuador); ERC PRG, TARISTU24-TK10 and MoER TK202 (Estonia); Academy of Finland, MEC, and HIP (Finland); CEA and CNRS/IN2P3 (France); SRNSF (Georgia); BMFTR, DFG, and HGF (Germany); GSRI (Greece); NKFIH (Hungary); DAE and DST (India); IPM (Iran); SFI (Ireland); INFN (Italy); MSIT and NRF (Republic of Korea); MES (Latvia); LMTLT (Lithuania); MOE and UM (Malaysia); BUAP, CINVESTAV, CONACYT, LNS, SEP, and UASLP-FAI (Mexico); MOS (Montenegro); MBIE (New Zealand); PAEC (Pakistan); MES, NSC, and NAWA (Poland); FCT (Portugal); MESTD (Serbia); MICIU/AEI and PCTI (Spain); MOSTR (Sri Lanka); Swiss Funding Agencies (Switzerland); MST (Taipei); MHESI (Thailand); TUBITAK and TENMAK (T\"{u}rkiye); NASU (Ukraine); STFC (United Kingdom); DOE and NSF (USA).

\hyphenation{Rachada-pisek} Individuals have received support from the Marie-Curie program and the European Research Council and Horizon 2020 Grant, contract Nos.\ 675440, 724704, 752730, 758316, 765710, 824093, 101115353, 101002207, 101001205, and COST Action CA16108 (European Union); the Leventis Foundation; the Alfred P.\ Sloan Foundation; the Alexander von Humboldt Foundation; the Science Committee, project no. 22rl-037 (Armenia); the Fonds pour la Formation \`a la Recherche dans l'Industrie et dans l'Agriculture (FRIA) and Fonds voor Wetenschappelijk Onderzoek contract No. 1228724N (Belgium); the Beijing Municipal Science \& Technology Commission, No. Z191100007219010, the Fundamental Research Funds for the Central Universities, the Ministry of Science and Technology of China under Grant No. 2023YFA1605804, the Natural Science Foundation of China under Grant No. 12535004, and USTC Research Funds of the Double First-Class Initiative No.\ YD2030002017 (China); the Ministry of Education, Youth and Sports (MEYS) of the Czech Republic; the Shota Rustaveli National Science Foundation, grant FR-22-985 (Georgia); the Deutsche Forschungsgemeinschaft (DFG), among others, under Germany's Excellence Strategy -- EXC 2121 ``Quantum Universe" -- 390833306, and under project number 400140256 - GRK2497; the Hellenic Foundation for Research and Innovation (HFRI), Project Number 2288 (Greece); the Hungarian Academy of Sciences, the New National Excellence Program - \'UNKP, the NKFIH research grants K 131991, K 133046, K 138136, K 143460, K 143477, K 146913, K 146914, K 147048, 2020-2.2.1-ED-2021-00181, TKP2021-NKTA-64, and 2025-1.1.5-NEMZ\_KI-2025-00004 (Hungary); the Council of Science and Industrial Research, India; ICSC -- National Research Center for High Performance Computing, Big Data and Quantum Computing, FAIR -- Future Artificial Intelligence Research, and CUP I53D23001070006 (Mission 4 Component 1), funded by the NextGenerationEU program (Italy); the Latvian Council of Science; the Ministry of Education and Science, project no. 2022/WK/14, and the National Science Center, contracts Opus 2021/41/B/ST2/01369, 2021/43/B/ST2/01552, 2023/49/B/ST2/03273, and the NAWA contract BPN/PPO/2021/1/00011 (Poland); the Funda\c{c}\~ao para a Ci\^encia e a Tecnologia, grant CEECIND/01334/2018 (Portugal); the National Priorities Research Program by Qatar National Research Fund; MICIU/AEI/10.13039/501100011033, ERDF/EU, ``European Union NextGenerationEU/PRTR", and Programa Severo Ochoa del Principado de Asturias (Spain); the Chulalongkorn Academic into Its 2nd Century Project Advancement Project, the National Science, Research and Innovation Fund program IND\_FF\_68\_369\_2300\_097, and the Program Management Unit for Human Resources \& Institutional Development, Research and Innovation, grant B39G680009 (Thailand); the Kavli Foundation; the Nvidia Corporation; the SuperMicro Corporation; the Welch Foundation, contract C-1845; and the Weston Havens Foundation (USA).
\end{acknowledgments}\section*{Data availability} Release and preservation of data used by the CMS Collaboration as the basis for publications is guided by the  \href{https://doi.org/10.7483/OPENDATA.CMS.1BNU.8V1W}{CMS data preservation, re-use and open access policy}.
\bibliography{auto_generated}

\providecommand{\href}[2]{#2}\begingroup\raggedright\begin{thebibliography}{10}%
\makeatletter
\providecommand{\hrefCMSnoop }[0]{\@secondoftwo}%
\makeatother
\providecommand{\doi}{\texttt{doi:}\begingroup \urlstyle{tt}\Url}

\bibitem{antgc_hagiwara}
\hrefCMSnoop {}{K.~Hagiwara, R.~D. Peccei, D.~Zeppenfeld, and K.~Hikasa,
  ``Probing the weak boson sector in {$\Pep+ \Pem \to \PWp + \PWm$}'',}
  \textit{ Nucl. Phys. B} \textbf{ 282} (1987) 253,
  \href{http://dx.doi.org/10.1016/0550-3213(87)90685-7}{\doi{10.1016/0550-3213(87)90685-7}}.

\bibitem{antgc_baur1}
\hrefCMSnoop {}{U.~Baur and E.~L. Berger, ``Probing the weak-boson sector in
  {$\PZ\gamma$} production at hadron colliders'',} \textit{ Phys. Rev. D}
  \textbf{ 47} (1993) 4889,
  \href{http://dx.doi.org/10.1103/PhysRevD.47.4889}{\doi{10.1103/PhysRevD.47.4889}}.

\bibitem{antgc_baur2}
\hrefCMSnoop {}{U.~Baur, T.~Han, and J.~Ohnemus, ``{QCD} corrections and
  anomalous couplings in {$\PZ\gamma$} production at hadron colliders'',}
  \textit{ Phys. Rev. D} \textbf{ 57} (1998) 2823,
  \href{http://dx.doi.org/10.1103/PhysRevD.57.2823}{\doi{10.1103/PhysRevD.57.2823}},
  \href{http://www.arXiv.org/abs/hep-ph/9710416}{\texttt{arXiv:hep-ph/9710416}}.

\bibitem{antgc_gounaris_1}
\hrefCMSnoop {}{G.~J. Gounaris, J.~Layssac, and F.~M. Renard, ``{Signatures of
  the anomalous $\PZ\gamma$ and $\mathrm{ZZ}$ production at lepton and hadron
  colliders}'',} \textit{ Phys. Rev. D} \textbf{ 61} (2000) 073013,
  \href{http://dx.doi.org/10.1103/PhysRevD.61.073013}{\doi{10.1103/PhysRevD.61.073013}},
  \href{http://www.arXiv.org/abs/hep-ph/9910395}{\texttt{arXiv:hep-ph/9910395}}.

\bibitem{antgcInterferenceBellazzini}
\hrefCMSnoop {}{B.~Bellazzini and F.~Riva, ``{New phenomenological and
  theoretical perspective on anomalous $ZZ$ and $Z\gamma$ processes}'',}
  \textit{ Phys. Rev. D} \textbf{ 98} (2018) 095021,
  \href{http://dx.doi.org/10.1103/PhysRevD.98.095021}{\doi{10.1103/PhysRevD.98.095021}},
  \href{http://www.arXiv.org/abs/1806.09640}{\texttt{arXiv:1806.09640}}.

\bibitem{Ellis:2022zdw}
\hrefCMSnoop {}{J.~Ellis, H.-J. He, and R.-Q. Xiao, ``{Probing neutral triple
  gauge couplings at the LHC and future hadron colliders}'',} \textit{ Phys.
  Rev. D} \textbf{ 107} (2023), no.~3, 035005,
  \href{http://dx.doi.org/10.1103/PhysRevD.107.035005}{\doi{10.1103/PhysRevD.107.035005}},
  \href{http://www.arXiv.org/abs/2206.11676}{\texttt{arXiv:2206.11676}}.

\bibitem{CMS:2025cxp}
\hrefCMSnoop {}{{CMS Collaboration}, ``{Measurement of Z$\gamma$ production in
  proton-proton collisions at $\sqrt{s}$ = 13.6 TeV and constraints on neutral
  triple gauge couplings}'',} 12, 2025.
  \href{http://www.arXiv.org/abs/2512.08582}{\texttt{arXiv:2512.08582}}.

\bibitem{d0_ZG_study_2005}
\hrefCMSnoop {}{{D0} Collaboration, ``Study of {$\PZ\gamma$} events and limits
  on anomalous {$\PZ\PZ\gamma$} and {$\PZ\gamma\gamma$} couplings in
  {$\Pp\PAp$} collisions at $\sqrt{s}=1.96$ {TeV}'',} \textit{ Phys. Rev.
  Lett.} \textbf{ 95} (2005) 051802,
  \href{http://dx.doi.org/10.1103/PhysRevLett.95.051802}{\doi{10.1103/PhysRevLett.95.051802}},
  \href{http://www.arXiv.org/abs/hep-ex/0502036}{\texttt{arXiv:hep-ex/0502036}}.

\bibitem{cdfZGmeasurement2010}
\hrefCMSnoop {}{{CDF} Collaboration, ``Measurement of {$\PZ\gamma$} production
  in {$\Pp\PAp$} collisions at {$\sqrt{s}=1.96$} {TeV}'',} \textit{ Phys. Rev.
  D} \textbf{ 82} (2010) 031103,
  \href{http://dx.doi.org/10.1103/PhysRevD.82.031103}{\doi{10.1103/PhysRevD.82.031103}},
  \href{http://www.arXiv.org/abs/1004.1140}{\texttt{arXiv:1004.1140}}.

\bibitem{cdfAntgcSearch2011}
\hrefCMSnoop {}{{CDF} Collaboration, ``{Limits on anomalous trilinear gauge
  couplings in $\PZ\PGg$ events from {$\Pp\PAp$} collisions at
  $\sqrt{s}=1.96$\TeV}'',} \textit{ Phys. Rev. Lett.} \textbf{ 107} (2011)
  051802,
  \href{http://dx.doi.org/10.1103/PhysRevLett.107.051802}{\doi{10.1103/PhysRevLett.107.051802}},
  \href{http://www.arXiv.org/abs/1103.2990}{\texttt{arXiv:1103.2990}}.

\bibitem{CMS_8TeV_2}
\hrefCMSnoop {}{{CMS Collaboration}, ``Measurement of the {$\PZ\gamma$}
  production cross section in pp collisions at {$\sqrt{s}=8$ TeV} and search
  for anomalous triple gauge boson couplings'',} \textit{ JHEP} \textbf{ 15}
  (2015) 164,
  \href{http://dx.doi.org/10.1007/JHEP04(2015)164}{\doi{10.1007/JHEP04(2015)164}},
  \href{http://www.arXiv.org/abs/1502.05664}{\texttt{arXiv:1502.05664}}.

\bibitem{CMS_8TeV_1}
\hrefCMSnoop {}{{CMS Collaboration}, ``Measurement of the {$\PZ\gamma
  \rightarrow \nu\bar{\nu}\gamma$} production cross section in ppcollisions at
  {$\sqrt{s}=8$ TeV} and limits on anomalous {$\PZ\PZ\gamma$} and
  {$\PZ\gamma\gamma$} trilinear gauge boson couplings'',} \textit{ Phys. Lett.
  B} \textbf{ 760} (2016) 448,
  \href{http://dx.doi.org/10.1016/j.physletb.2016.06.080}{\doi{10.1016/j.physletb.2016.06.080}},
  \href{http://www.arXiv.org/abs/1602.07152}{\texttt{arXiv:1602.07152}}.

\bibitem{atlas_8TeV}
\hrefCMSnoop {}{{ATLAS Collaboration}, ``Measurements of {$\PZ\gamma$} and
  {$\PZ\gamma\gamma$} production in pp collisions at {$\sqrt{s}=8$ TeV} with
  the {ATLAS} detector'',} \textit{ Phys. Rev. D} \textbf{ 93} (2016) 112002,
  \href{http://dx.doi.org/10.1103/PhysRevD.93.112002}{\doi{10.1103/PhysRevD.93.112002}},
  \href{http://www.arXiv.org/abs/1604.05232}{\texttt{arXiv:1604.05232}}.

\bibitem{atlas_13TeV}
\hrefCMSnoop {}{{ATLAS Collaboration}, ``Measurement of the {$\PZ\gamma
  \rightarrow \nu\bar{\nu}\gamma$} production cross section in {pp} collisions
  at {$\sqrt{s}=13$ TeV} with the {ATLAS} detector and limits on anomalous
  triple gauge-boson couplings'',} \textit{ JHEP} \textbf{ 12} (2018) 010,
  \href{http://dx.doi.org/10.1007/JHEP12(2018)010}{\doi{10.1007/JHEP12(2018)010}},
  \href{http://www.arXiv.org/abs/1810.04995}{\texttt{arXiv:1810.04995}}.

\bibitem{d0_ZG_llg_study_2007}
\hrefCMSnoop {}{{D0} Collaboration, ``{$\PZ\gamma$} production and limits on
  anomalous {$\PZ\PZ\gamma$} and {$\PZ\gamma\gamma$} couplings in {$\Pp\PAp$}
  collisions at {$\sqrt{s}=1.96$ TeV}'',} \textit{ Phys. Lett. B} \textbf{ 653}
  (2007) 378,
  \href{http://dx.doi.org/10.1016/j.physletb.2007.08.035}{\doi{10.1016/j.physletb.2007.08.035}},
  \href{http://www.arXiv.org/abs/0705.1550}{\texttt{arXiv:0705.1550}}.

\bibitem{atlas_7tev_llg}
\hrefCMSnoop {}{{The ATLAS collaboration}, ``{Measurement of W$\gamma$ and
  Z$\gamma$ production in proton-proton collisions at $\sqrt{s}=7$ TeV with the
  ATLAS detector}'',} \textit{ JHEP} \textbf{ 09} (2011) 072,
  \href{http://dx.doi.org/10.1007/JHEP09(2011)072}{\doi{10.1007/JHEP09(2011)072}},
  \href{http://www.arXiv.org/abs/1106.1592}{\texttt{arXiv:1106.1592}}.

\bibitem{hepdata}
\hrefCMSnoop {}{}{HEPData} record for this analysis, 2026.
\newblock
  \href{http://dx.doi.org/10.17182/hepdata.167736}{\doi{10.17182/hepdata.167736}}.

\bibitem{CMS:2008xjf}
\hrefCMSnoop {}{{CMS Collaboration}, ``The {CMS} experiment at the {CERN}
  {LHC}'',} \textit{ JINST} \textbf{ 3} (2008) S08004,
  \href{http://dx.doi.org/10.1088/1748-0221/3/08/S08004}{\doi{10.1088/1748-0221/3/08/S08004}}.

\bibitem{CMS:2023gfb}
\hrefCMSnoop {}{{CMS Collaboration}, ``Development of the {CMS} detector for
  the {CERN LHC Run 3}'',} \textit{ JINST} \textbf{ 19} (2024) P05064,
  \href{http://dx.doi.org/10.1088/1748-0221/19/05/P05064}{\doi{10.1088/1748-0221/19/05/P05064}},
  \href{http://www.arXiv.org/abs/2309.05466}{\texttt{arXiv:2309.05466}}.

\bibitem{CMS:2020cmk}
\hrefCMSnoop {}{{CMS Collaboration}, ``{Performance of the CMS Level-1 trigger
  in proton-proton collisions at $\sqrt{s} = 13${TeV}}'',} \textit{ JINST}
  \textbf{ 15} (2020) P10017,
  \href{http://dx.doi.org/10.1088/1748-0221/15/10/P10017}{\doi{10.1088/1748-0221/15/10/P10017}},
  \href{http://www.arXiv.org/abs/2006.10165}{\texttt{arXiv:2006.10165}}.

\bibitem{CMS:2016ngn}
\hrefCMSnoop {}{{CMS Collaboration}, ``The {CMS} trigger system'',} \textit{
  JINST} \textbf{ 12} (2017) P01020,
  \href{http://dx.doi.org/10.1088/1748-0221/12/01/P01020}{\doi{10.1088/1748-0221/12/01/P01020}},
\href{http://www.arXiv.org/abs/1609.02366}{\texttt{arXiv:1609.02366}}.

\bibitem{CMS:2024aqx}
\hrefCMSnoop {}{{CMS Collaboration}, ``Performance of the {CMS} high-level
  trigger during {LHC} run 2'',} \textit{ JINST} \textbf{ 19} (2024) P11021,
  \href{http://dx.doi.org/10.1088/1748-0221/19/11/P11021}{\doi{10.1088/1748-0221/19/11/P11021}},
  \href{http://www.arXiv.org/abs/2410.17038}{\texttt{arXiv:2410.17038}}.

\bibitem{CMS:2020uim}
\hrefCMSnoop {}{{CMS Collaboration}, ``{Electron and photon reconstruction and
  identification with the CMS experiment at the CERN LHC}'',} \textit{ JINST}
  \textbf{ 16} (2021) P05014,
  \href{http://dx.doi.org/10.1088/1748-0221/16/05/P05014}{\doi{10.1088/1748-0221/16/05/P05014}},
  \href{http://www.arXiv.org/abs/2012.06888}{\texttt{arXiv:2012.06888}}.

\bibitem{CMS:2018rym}
\hrefCMSnoop {}{{CMS Collaboration}, ``Performance of the {CMS} muon detector
  and muon reconstruction with proton-proton collisions at
  $\sqrt{s}=13$\,{TeV}'',} \textit{ JINST} \textbf{ 13} (2018) P06015,
  \href{http://dx.doi.org/10.1088/1748-0221/13/06/P06015}{\doi{10.1088/1748-0221/13/06/P06015}},
  \href{http://www.arXiv.org/abs/1804.04528}{\texttt{arXiv:1804.04528}}.

\bibitem{CMS:2014pgm}
\hrefCMSnoop {}{{CMS Collaboration}, ``Description and performance of track and
  primary-vertex reconstruction with the {CMS} tracker'',} \textit{ JINST}
  \textbf{ 9} (2014) P10009,
  \href{http://dx.doi.org/10.1088/1748-0221/9/10/P10009}{\doi{10.1088/1748-0221/9/10/P10009}},
  \href{http://www.arXiv.org/abs/1405.6569}{\texttt{arXiv:1405.6569}}.

\bibitem{CMS:2017yfk}
\hrefCMSnoop {}{{CMS Collaboration}, ``Particle-flow reconstruction and global
  event description with the {CMS} detector'',} \textit{ JINST} \textbf{ 12}
  (2017) P10003,
  \href{http://dx.doi.org/10.1088/1748-0221/12/10/P10003}{\doi{10.1088/1748-0221/12/10/P10003}},
\href{http://www.arXiv.org/abs/1706.04965}{\texttt{arXiv:1706.04965}}.

\bibitem{CMS-TDR-15-02}
\href {http://cds.cern.ch/record/2020886}{{CMS Collaboration}, ``Technical
  proposal for the {Phase-II} upgrade of the {Compact Muon Solenoid}'',} CMS
  Technical Proposal CERN-LHCC-2015-010, CMS-TDR-15-02, 2015.

\bibitem{Cacciari:2008gp}
\hrefCMSnoop {}{M.~Cacciari, G.~P. Salam, and G.~Soyez, ``The anti-\kt jet
  clustering algorithm'',} \textit{ JHEP} \textbf{ 04} (2008) 063,
  \href{http://dx.doi.org/10.1088/1126-6708/2008/04/063}{\doi{10.1088/1126-6708/2008/04/063}},
  \href{http://www.arXiv.org/abs/0802.1189}{\texttt{arXiv:0802.1189}}.

\bibitem{Cacciari:2011ma}
\hrefCMSnoop {}{M.~Cacciari, G.~P. Salam, and G.~Soyez, ``{FastJet} user
  manual'',} \textit{ Eur. Phys. J. C} \textbf{ 72} (2012) 1896,
  \href{http://dx.doi.org/10.1140/epjc/s10052-012-1896-2}{\doi{10.1140/epjc/s10052-012-1896-2}},
\href{http://www.arXiv.org/abs/1111.6097}{\texttt{arXiv:1111.6097}}.

\bibitem{Sirunyan:2020foa}
\hrefCMSnoop {}{{CMS Collaboration}, ``{Pileup mitigation at CMS in 13 TeV
  data}'',} \textit{ JINST} \textbf{ 15} (2020) P09018,
  \href{http://dx.doi.org/10.1088/1748-0221/15/09/p09018}{\doi{10.1088/1748-0221/15/09/p09018}},
  \href{http://www.arXiv.org/abs/2003.00503}{\texttt{arXiv:2003.00503}}.

\bibitem{CMS:2016lmd}
\hrefCMSnoop {}{{CMS Collaboration}, ``Jet energy scale and resolution in the
  {CMS} experiment in pp collisions at 8 {TeV}'',} \textit{ JINST} \textbf{ 12}
  (2017) P02014,
  \href{http://dx.doi.org/10.1088/1748-0221/12/02/P02014}{\doi{10.1088/1748-0221/12/02/P02014}},
\href{http://www.arXiv.org/abs/1607.03663}{\texttt{arXiv:1607.03663}}.

\bibitem{Sirunyan:2019kia}
\hrefCMSnoop {}{{CMS Collaboration}, ``Performance of missing transverse
  momentum reconstruction in proton-proton collisions at $\sqrt{s} = 13$\,{TeV}
  using the {CMS} detector'',} \textit{ JINST} \textbf{ 14} (2019) P07004,
  \href{http://dx.doi.org/10.1088/1748-0221/14/07/P07004}{\doi{10.1088/1748-0221/14/07/P07004}},
\href{http://www.arXiv.org/abs/1903.06078}{\texttt{arXiv:1903.06078}}.

\bibitem{Khachatryan:2010xn}
\hrefCMSnoop {}{{CMS Collaboration}, ``Measurements of inclusive {$\PW$} and
  {$\PZ$} cross sections in {$\Pp\Pp$} collisions at $\sqrt{s}=7$ {TeV}'',}
  \textit{ JHEP} \textbf{ 01} (2011) 080,
  \href{http://dx.doi.org/10.1007/JHEP01(2011)080}{\doi{10.1007/JHEP01(2011)080}},
  \href{http://www.arXiv.org/abs/1012.2466}{\texttt{arXiv:1012.2466}}.

\bibitem{Bodek:2012id}
A.~Bodek\hrefCMSnoop {}{ { et~al.}, ``Extracting muon momentum scale
  corrections for hadron collider experiments'',} \textit{ Eur. Phys. J. C}
  \textbf{ 72} (2012) 2194,
  \href{http://dx.doi.org/10.1140/epjc/s10052-012-2194-8}{\doi{10.1140/epjc/s10052-012-2194-8}},
\href{http://www.arXiv.org/abs/1208.3710}{\texttt{arXiv:1208.3710}}.

\bibitem{spikes_petyt_mitigation_2012}
\hrefCMSnoop {}{{CMS Collaboration}, ``{Mitigation of anomalous APD signals in
  the CMS electromagnetic calorimeter}'',} \textit{ J. Phys. Conf. Ser.}
  \textbf{ 404} (2012) 012043,
  \href{http://dx.doi.org/10.1088/1742-6596/404/1/012043}{\doi{10.1088/1742-6596/404/1/012043}}.

\bibitem{beamHaloSingh2010}
\hrefCMSnoop {}{A.~P. Singh, P.~C. Bhat, N.~V. Mokhov, and S.~Beri,
  ``Beam-induced radiation in the compact muon solenoid tracker at the large
  hadron collider'',} \textit{ Pramana} \textbf{ 74} (2010) 719,
  \href{http://dx.doi.org/10.1007/s12043-010-0093-9}{\doi{10.1007/s12043-010-0093-9}}.

\bibitem{MadGraph}
J.~Alwall\hrefCMSnoop {}{ { et~al.}, ``The automated computation of tree-level
  and next-to-leading order differential cross sections, and their matching to
  parton shower simulations'',} \textit{ JHEP} \textbf{ 07} (2014) 079,
  \href{http://dx.doi.org/10.1007/JHEP07(2014)079}{\doi{10.1007/JHEP07(2014)079}},
  \href{http://www.arXiv.org/abs/1405.0301}{\texttt{arXiv:1405.0301}}.

\bibitem{Sirunyan:2019dfx}
\hrefCMSnoop {}{{CMS Collaboration}, ``Extraction and validation of a new set
  of {CMS} {PYTHIA8} tunes from underlying-event measurements'',} \textit{ Eur.
  Phys. J. C} \textbf{ 80} (2020) 4,
  \href{http://dx.doi.org/10.1140/epjc/s10052-019-7499-4}{\doi{10.1140/epjc/s10052-019-7499-4}},
  \href{http://www.arXiv.org/abs/1903.12179}{\texttt{arXiv:1903.12179}}.

\bibitem{pythia8}
T.~Sj{\"o}strand\hrefCMSnoop {}{ { et~al.}, ``An introduction to {PYTHIA
  8.2}'',} \textit{ Comput. Phys. Commun.} \textbf{ 191} (2015) 159,
  \href{http://dx.doi.org/10.1016/j.cpc.2015.01.024}{\doi{10.1016/j.cpc.2015.01.024}},
  \href{http://www.arXiv.org/abs/1410.3012}{\texttt{arXiv:1410.3012}}.

\bibitem{GEANT4}
\hrefCMSnoop {}{{GEANT4} Collaboration, ``{\GEANTfour}---a simulation
  toolkit'',} \textit{ Nucl. Instrum. Meth. A} \textbf{ 506} (2003) 250,
  \href{http://dx.doi.org/10.1016/S0168-9002(03)01368-8}{\doi{10.1016/S0168-9002(03)01368-8}}.

\bibitem{kFacQCD_Grazzini}
\hrefCMSnoop {}{M.~Grazzini, S.~Kallweit, and D.~Rathlev, ``{W$\gamma$ and
  Z$\gamma$ production at the LHC in NNLO QCD}'',} \textit{ JHEP} \textbf{ 07}
  (2015) 085,
  \href{http://dx.doi.org/10.1007/JHEP07(2015)085}{\doi{10.1007/JHEP07(2015)085}},
  \href{http://www.arXiv.org/abs/1504.01330}{\texttt{arXiv:1504.01330}}.

\bibitem{kFacEWZG_Denner}
\hrefCMSnoop {}{A.~Denner, S.~Dittmaier, M.~Hecht, and C.~Pasold, ``{NLO QCD
  and electroweak corrections to Z + $\gamma$ production with leptonic Z-boson
  decays}'',} \textit{ JHEP} \textbf{ 02} (2016) 057,
  \href{http://dx.doi.org/10.1007/JHEP02(2016)057}{\doi{10.1007/JHEP02(2016)057}},
  \href{http://www.arXiv.org/abs/1510.08742}{\texttt{arXiv:1510.08742}}.

\bibitem{kFacEWWG_Denner}
\hrefCMSnoop {}{A.~Denner, S.~Dittmaier, M.~Hecht, and C.~Pasold, ``{NLO QCD
  and electroweak corrections to W + $\gamma$ production with leptonic W-boson
  decays}'',} \textit{ JHEP} \textbf{ 04} (2015) 018,
  \href{http://dx.doi.org/10.1007/JHEP04(2015)018}{\doi{10.1007/JHEP04(2015)018}},
  \href{http://www.arXiv.org/abs/1412.7421}{\texttt{arXiv:1412.7421}}.

\bibitem{Bertone:2017bme}
\hrefCMSnoop {}{V.~Bertone, S.~Carrazza, N.~P. Hartland, and J.~Rojo,
  ``{Illuminating the photon content of the proton within a global PDF
  analysis}'',} \textit{ SciPost Phys.} \textbf{ 5} (2018) 008,
  \href{http://dx.doi.org/10.21468/SciPostPhys.5.1.008}{\doi{10.21468/SciPostPhys.5.1.008}},
  \href{http://www.arXiv.org/abs/1712.07053}{\texttt{arXiv:1712.07053}}.

\bibitem{WadudMohammadAbrar2023Motp}
\href {https://cds.cern.ch/record/2883218}{M.~A. Wadud, ``{Measurements on the
  production of a Z boson in association with a photon in proton-proton
  collisions and a search for anomalous triple gauge couplings at the CERN
  Large Hadron Collider}''}.
\newblock PhD thesis, University of Minnesota, 2023.

\bibitem{CMS-DP-2020-040}
\href {https://cds.cern.ch/record/2727091}{{CMS Collaboration}, ``{Muon
  reconstruction and identification performance with Run-2 data}'',} CMS
  Detector Performance Summary CMS-DP-2020-040, 2020.

\bibitem{cms_dm_search_2019}
\hrefCMSnoop {}{{CMS collaboration}, ``Search for new physics in final states
  with a single photon and missing transverse momentum in proton-proton
  collisions at {$\sqrt{s}=13$} {TeV}'',} \textit{ JHEP} \textbf{ 02} (Feb,
  2019) 074,
  \href{http://dx.doi.org/10.1007/JHEP02(2019)074}{\doi{10.1007/JHEP02(2019)074}},
  \href{http://www.arXiv.org/abs/1810.00196}{\texttt{arXiv:1810.00196}}.

\bibitem{lumi2016}
\hrefCMSnoop {}{{CMS Collaboration}, ``{Precision luminosity measurement in
  proton-proton collisions at $\sqrt{s} =$ 13 TeV in 2015 and 2016 at CMS}'',}
  \textit{ Eur. Phys. J. C} \textbf{ 81} (2021), no.~9, 800,
  \href{http://dx.doi.org/10.1140/epjc/s10052-021-09538-2}{\doi{10.1140/epjc/s10052-021-09538-2}},
  \href{http://www.arXiv.org/abs/2104.01927}{\texttt{arXiv:2104.01927}}.

\bibitem{lumi2017}
\href {http://cds.cern.ch/record/2621960}{{CMS Collaboration}, ``{CMS}
  luminosity measurement for the 2017 data-taking period at
  $\sqrt{s}={13\TeV}$'',} CMS Physics Analysis Summary CMS-PAS-LUM-17-004,
  2018.

\bibitem{lumi2018}
\href {http://cds.cern.ch/record/2676164}{{CMS Collaboration}, ``{CMS}
  luminosity measurement for the 2018 data-taking period at
  $\sqrt{s}={13\TeV}$'',} CMS Physics Analysis Summary CMS-PAS-LUM-18-002,
  2019.

\bibitem{CMS:2018mlc}
\hrefCMSnoop {}{{CMS Collaboration}, ``{Measurement of the inelastic
  proton-proton cross section at $ \sqrt{s}=13 $ TeV}'',} \textit{ JHEP}
  \textbf{ 07} (2018) 161,
  \href{http://dx.doi.org/10.1007/JHEP07(2018)161}{\doi{10.1007/JHEP07(2018)161}},
  \href{http://www.arXiv.org/abs/1802.02613}{\texttt{arXiv:1802.02613}}.

\bibitem{matrix_Grazzini2018}
\hrefCMSnoop {}{M.~Grazzini, S.~Kallweit, and M.~Wiesemann, ``Fully
  differential {NNLO} computations with {MATRIX}'',} \textit{ Eur. Phys. J. C}
  \textbf{ 78} (2018) 537,
  \href{http://dx.doi.org/10.1140/epjc/s10052-018-5771-7}{\doi{10.1140/epjc/s10052-018-5771-7}},
  \href{http://www.arXiv.org/abs/1711.06631}{\texttt{arXiv:1711.06631}}.

\bibitem{nnloQcdUncLindert2017}
\hrefCMSnoop {}{J.~M. Lindert { et~al.}, ``{Precise predictions for $V+$ jets
  dark matter backgrounds}'',} \textit{ Eur. Phys. J. C} \textbf{ 77} (2017),
  no.~12, 829,
  \href{http://dx.doi.org/10.1140/epjc/s10052-017-5389-1}{\doi{10.1140/epjc/s10052-017-5389-1}},
  \href{http://www.arXiv.org/abs/1705.04664}{\texttt{arXiv:1705.04664}}.

\bibitem{butterworth_pdf4lhc_2016}
\hrefCMSnoop {}{J.~Butterworth { et~al.}, ``{PDF4LHC recommendations for LHC
  Run II}'',} \textit{ J. Phys. G} \textbf{ 43} (2016) 023001,
  \href{http://dx.doi.org/10.1088/0954-3899/43/2/023001}{\doi{10.1088/0954-3899/43/2/023001}},
  \href{http://www.arXiv.org/abs/1510.03865}{\texttt{arXiv:1510.03865}}.

\bibitem{CMS:2024onh}
\hrefCMSnoop {}{{CMS Collaboration}, ``{The CMS statistical analysis and
  combination tool: Combine}'',} \textit{ Comput. Softw. Big Sci.} \textbf{ 8}
  (2024), no.~1, 19,
  \href{http://dx.doi.org/10.1007/s41781-024-00121-4}{\doi{10.1007/s41781-024-00121-4}},
  \href{http://www.arXiv.org/abs/2404.06614}{\texttt{arXiv:2404.06614}}.

\bibitem{Barlow:1993dm}
\hrefCMSnoop {}{R.~J. Barlow and C.~Beeston, ``{Fitting using finite Monte
  Carlo samples}'',} \textit{ Comput. Phys. Commun.} \textbf{ 77} (1993) 219,
  \href{http://dx.doi.org/10.1016/0010-4655(93)90005-W}{\doi{10.1016/0010-4655(93)90005-W}}.

\bibitem{unfolding_GAPONENKO2020163612}
\hrefCMSnoop {}{A.~Gaponenko, ``A practical way to regularize unfolding of
  sharply varying spectra with low data statistics'',} \textit{ Nucl. Instrum.
  Meth. A} \textbf{ 960} (2020) 163612,
  \href{http://dx.doi.org/10.1016/j.nima.2020.163612}{\doi{10.1016/j.nima.2020.163612}},
  \href{http://www.arXiv.org/abs/1906.07918}{\texttt{arXiv:1906.07918}}.

\bibitem{Pollack:2017srh}
\hrefCMSnoop {}{B.~Pollack, S.~Bhattacharya, and M.~Schmitt, ``{Bayesian block
  histogramming for high energy physics}'',} 2017.
  \href{http://www.arXiv.org/abs/1708.00810}{\texttt{arXiv:1708.00810}}.

\end{thebibliography}\endgroup
\cleardoublepage \appendix\section{The CMS Collaboration \label{app:collab}}\begin{sloppypar}\hyphenpenalty=5000\widowpenalty=500\clubpenalty=5000\cmsinstitute{Yerevan Physics Institute, Yerevan, Armenia}
{\tolerance=6000
A.~Hayrapetyan, A.~Tumasyan\cmsAuthorMark{1}\cmsorcid{0009-0000-0684-6742}
\par}
\cmsinstitute{Institut f\"{u}r Hochenergiephysik, Vienna, Austria}
{\tolerance=6000
W.~Adam\cmsorcid{0000-0001-9099-4341}, J.W.~Andrejkovic, T.~Bergauer\cmsorcid{0000-0002-5786-0293}, S.~Chatterjee\cmsorcid{0000-0003-2660-0349}, K.~Damanakis\cmsorcid{0000-0001-5389-2872}, M.~Dragicevic\cmsorcid{0000-0003-1967-6783}, P.S.~Hussain\cmsorcid{0000-0002-4825-5278}, M.~Jeitler\cmsAuthorMark{2}\cmsorcid{0000-0002-5141-9560}, N.~Krammer\cmsorcid{0000-0002-0548-0985}, A.~Li\cmsorcid{0000-0002-4547-116X}, D.~Liko\cmsorcid{0000-0002-3380-473X}, I.~Mikulec\cmsorcid{0000-0003-0385-2746}, J.~Schieck\cmsAuthorMark{2}\cmsorcid{0000-0002-1058-8093}, D.~Schwarz\cmsorcid{0000-0002-3821-7331}, R.~Sch\"{o}fbeck\cmsAuthorMark{2}\cmsorcid{0000-0002-2332-8784}, M.~Sonawane\cmsorcid{0000-0003-0510-7010}, W.~Waltenberger\cmsorcid{0000-0002-6215-7228}, C.E.~Wulz\cmsAuthorMark{2}\cmsorcid{0000-0001-9226-5812}
% ...
\par}
$^{1}$Also at Yerevan State University, Yerevan, Armenia\\
$^{2}$Also at Technische Universit\"{a}t Wien, Vienna, Austria\\
% ...

\end{sloppypar}
\end{document}